\documentclass[12pt]{article}
\usepackage[dvips]{graphicx}
\usepackage[small]{caption}

\usepackage{epsfig,amsmath,amssymb,verbatim,mathrsfs}
\usepackage[square, comma, sort&compress]{natbib}
\usepackage{hyperref}
\numberwithin{equation}{section}

\def\beq{\begin{eqnarray}}
\def\eeq{\end{eqnarray}}
\def\bea{\begin{eqnarray}}
\def\eea{\end{eqnarray}}

\def\gev{\, {\rm GeV}}

\newcommand{\gsim}{\lower.7ex\hbox{$\;\stackrel{\textstyle>}{\sim}\;$}}
\newcommand{\lsim}{\lower.7ex\hbox{$\;\stackrel{\textstyle<}{\sim}\;$}}
\def\trivrep{{\bf 1}}
\def\tenrep{{\bf 10}}
\def\tenbarrep{{\overline{\bf 10}}}
\def\fiverep{{\bf 5}}
\def\fivebarrep{\overline{\bf 5}}

\newcommand{\Eq}[1]{Eq.\;(\ref{#1})}
\newcommand{\Eqs}[1]{Eqs.\;(\ref{#1})}

\newcommand\SU{SU}
\newcommand\SO{SO}
\newcommand\Sp{Sp}
\newcommand\U{U}

\newcommand\GUT{\mathrm{GUT}}
\newcommand\pl{\mathrm{pl}}
\newcommand\Tr{\mathrm{Tr}}

\newcommand\x{\times}
\newcommand\cO{\mathcal{O}}

\newcommand\g{\gamma}
\newcommand\e{\epsilon}
\newcommand\s{\sigma}
\renewcommand\a{\alpha}
\renewcommand\th{\theta}
\newcommand\de{\delta}
\newcommand\cN{\mathcal{N}}
\newcommand\half{\frac 1 2}
\renewcommand\l{\lambda}
\renewcommand\L{\Lambda}
\renewcommand\b{\beta}
\newcommand\p[1]{\left(#1\right)}
\newcommand\ptl{\partial}
\newcommand\<{\langle}
\renewcommand\>{\rangle}
\renewcommand\bar[1]{\overline{#1}}
\renewcommand\hat[1]{\widehat{#1}}
\newcommand\tl[1]{\widetilde{#1}}
\newcommand\spceq{\ \ =\ \ }
\renewcommand\.{\cdot}
\newcommand\tr{\mathrm{tr}}

\newcommand{\drawsquare}[2]{\hbox{%
\rule{#2pt}{#1pt}\hskip-#2pt%  left vertical
\rule{#1pt}{#2pt}\hskip-#1pt%  lower horizontal
\rule[#1pt]{#1pt}{#2pt}}\rule[#1pt]{#2pt}{#2pt}\hskip-#2pt%  upper horizontal
\rule{#2pt}{#1pt}}% right vertical

\newcommand{\Yfund}{\raisebox{-.5pt}{\drawsquare{6.5}{0.4}}}%  fund
\newcommand{\Yafund}{\overline{\Yfund}}%  anti-fund
\newcommand{\Ysymm}{\raisebox{-.5pt}{\drawsquare{6.5}{0.4}}\hskip-0.4pt%
        \raisebox{-.5pt}{\drawsquare{6.5}{0.4}}}%  symmetric second rank
\newcommand{\Yasymm}{\raisebox{-3.5pt}{\drawsquare{6.5}{0.4}}\hskip-6.9pt%
        \raisebox{3pt}{\drawsquare{6.5}{0.4}}}%  antisymmetric second rank

\begin{document}

\setlength{\baselineskip}{0.2in}

\begin{titlepage}
\noindent
\begin{flushright}
%HUTP-08-xx\\
\end{flushright}
\vspace{1cm}

\begin{center}
  \begin{Large}
    \begin{bf}
Superconformal Flavor Simplified\\
     \end{bf}
  \end{Large}
\end{center}
\vspace{0.2cm}

\begin{center}

\begin{large}
David Poland and David Simmons-Duffin\\
\end{large}
\vspace{0.3cm}
\begin{it}
Jefferson Physical Laboratory, Harvard University,\\
Cambridge, Massachusetts 02138, USA\\
\vspace{0.5cm}
\end{it}

\end{center}

\center{\today}

\begin{abstract}
A simple explanation of the flavor hierarchies can arise if matter
fields interact with a conformal sector and different generations
have different anomalous dimensions under the CFT.  However, in
the original study by Nelson and Strassler many supersymmetric
models of this type were considered to be `incalculable' because the
$R$-charges were not sufficiently constrained by the superpotential.
We point out that nearly all such models are calculable with the
use of $a$-maximization.  Utilizing this, we construct the simplest
vector-like flavor models and discuss their viability.
A significant constraint on these models comes from requiring
that the visible gauge couplings remain perturbative throughout the
conformal window needed to generate the hierarchies.  However, we find that
there is a small class of simple flavor models that can evade this bound.

\end{abstract}

\vspace{1cm}

\end{titlepage}

\setcounter{page}{2}

\tableofcontents

\vfill\eject

%%%%%%%%%%%%%%%%%%%%%%%%%%%%%%%%%%%%%%%%%%%%%%%%%%%%%%%%%%%%%%%%%%%%%%

\newpage

\section{Introduction}
\label{sec:intro}

While the Standard Model is highly successful as an effective field
theory in describing nature, one of its big mysteries is the observed
pattern of quark and lepton masses and mixing angles.  This puzzle
remains even after new physics such as supersymmetry or strong
dynamics is introduced in order to solve the hierarchy problem.  In
fact, there is quite often a tension between solutions to the
hierarchy problem and a successful picture of flavor physics, leading
to many models that, while quite interesting, perhaps seem more
contrived than what nature should allow.

A simple explanation of the structure of flavor physics can come about
if the Yukawa interactions are anarchical in the UV but each field
comes with an additional suppression factor in the IR effective
theory~\cite{Froggatt:1978nt,Leurer:1992wg,Leurer:1993gy,Grossman:1995hk}.
These suppression factors can arise in 4D theories if each field
carries a different anomalous dimension under strong conformal
dynamics~\cite{Georgi:1983mq,Nelson:2000sn}, or in warped 5D
theories~\cite{Randall:1999ee} if each field has a different
exponential profile as determined by its bulk
mass~\cite{Gherghetta:2000qt,Huber:2000ie}.  In fact, these two
pictures can be related via the AdS/CFT
correspondence~\cite{Maldacena:1997re,Gubser:1998bc,Witten:1998qj,ArkaniHamed:2000ds,Rattazzi:2000hs}.
The bulk masses in 5D theories are usually taken to be free
parameters, leading to a successful explanation of flavor physics, but
leaving open the question of whether there exists a dual 4D CFT
containing operators with the corresponding anomalous dimensions.

On the other hand, one can attempt to find concrete 4D theories in
which the anomalous dimensions can be determined as output, rather
than input.  Some examples of this in a supersymmetric context were
given in~\cite{Nelson:2000sn}.  However, these models were necessarily
somewhat complicated because, unless there were a large number of
marginal interactions in the superpotential, there was no general
method available to determine the anomalous dimensions (or
equivalently the $R$-charges) of every field.

Here we point out that $a$-maximization~\cite{Intriligator:2003jj} is
precisely the needed method, because it allows one to determine the
correct superconformal $U(1)_R$-symmetry even when there are only a
small number of additional constraints.  We will demonstrate that this
allows one to construct very simple calculable models of flavor physics,
as well as reanalyze the `incalculable' models presented in \cite{Nelson:2000sn}.

However, since these models inevitably introduce many new multiplets
charged under $\SU(3)\times \SU(2)\times \U(1)$, this flavor physics
must occur at a high scale in order for the visible gauge couplings
to remain perturbative.  In addition, many of these models introduce
couplings that violate baryon and lepton number, potentially inducing
problematic proton-decay operators.  A simple way to avoid both of
these constraints is to assume that the conformal dynamics occurs
above the GUT scale, and thus we focus on on finding the simplest
vector-like flavor models in the context of $\SU(5)_\GUT$ unification.
We will find that, even in this case, a significant constraint comes from
requiring that the $\SU(5)_\GUT$ gauge coupling remain perturbative
throughout the entire conformal window.  However, we will demonstrate
that a small class of models can also satisfy this constraint while
explaining the observed flavor hierarchies.

The outline of this work is as follows.  In Section~\ref{sec:overview}
we will give brief reviews of flavor physics and $a$-maximization.  In
Section~\ref{sec:sunmodels} we study simple concrete flavor models
with vector-like matter content.  In Section~\ref{sec:incalculable} we
analyze one of the previously `incalculable' models presented by Nelson
and Strassler.  We conclude in Section~\ref{sec:concl}.

\section{Overview}
\label{sec:overview}

\subsection{Review of Flavor Physics}
\label{sec:flavor}

We will start with a brief review of how the flavor hierarchies appear
in supersymmetric extensions of the Standard Model.  The Yukawa
couplings of the matter fields $Q_i,U_i,D_i,L_i,E_i$ to the Higgs
fields $H_u,H_d$ arise from the superpotential operators
\beq
W = y_u^{i j} Q_i U_j H_u + y_d^{i j} Q_i D_j H_d + y_l^{i j} L_i E_j H_d \,\left(\,+\, y_N^{i j} L_i N_j H_u + M^{i j}_N N_i N_j\,\right),
\eeq
where the matrices $y_a^{i j}$ possess a hierarchical structure in the
basis where all of the matter fields have a canonically normalized
K\"{a}hler potential.  (Here we have included the possibility of
generating neutrino masses via the seesaw mechanism by integrating out
massive right-handed neutrinos $N_i$.)

There are two fundamentally different ways that this hierarchical
structure could arise from an underlying flavor physics model.  The
first is that the model genuinely generates a hierarchical structure
in the superpotential while not significantly affecting the K\"{a}hler
potential.  This could happen, e.g., in models in which the couplings
of the lighter generations ultimately arise from higher dimensional
operators in the superpotential~\cite{Leurer:1992wg,Leurer:1993gy,Grossman:1995hk}.
In this case one would need to know the details of the model in order to
determine if there are any predicted relations between the masses and
mixing angles.

The second possibility is that the superpotential contains no
hierarchical structure (``flavor anarchy''), but the underlying
flavor physics model generates a hierarchy in the wave-function
factors \beq \mathcal{L} = \int d^4 \theta \sum_i Z_{i}
\Phi^{\dagger}_i \Phi^{\phantom{\dagger}}_i, \eeq with, e.g.,
$Z_{Q_1} \gg Z_{Q_2} \gg Z_{Q_3}$.  Taking $\epsilon_i \equiv
Z_i^{-1/2}$, then up to $O(1)$ coefficients one obtains the
generic mass structure \bea \left(m_t,\ m_c,\ m_u\right) &\approx&
\< H_u \> \left(\e_{Q_3} \e_{U_3} \e_{H_u} , \ \e_{Q_2} \e_{U_2}
\e_{H_u}, \
\e_{Q_1} \e_{U_1} \e_{H_u} \right) \nonumber\\
\left(m_b,\ m_s,\ m_d\right) &\approx&
\< H_d \>
\left(\e_{Q_3} \e_{D_3} \e_{H_d} , \
\e_{Q_2} \e_{D_2} \e_{H_d}, \
\e_{Q_1} \e_{D_1} \e_{H_d} \right) \nonumber\\
\left(m_{\tau},\ m_{\mu},\ m_e\right) &\approx&
\< H_d \>
\left(\e_{L_3} \e_{E_3} \e_{H_d} ,\
\e_{L_2} \e_{E_2} \e_{H_d},\
\e_{L_1} \e_{E_1} \e_{H_d} \right) \nonumber\\
\left(m_{\nu_{\tau}},\ m_{\nu_{\mu}},\ m_{\nu_e}\right) &\approx&
\frac{\< H_u \>^2 }{M_N} \left(\e_{L_3}^2
\e_{H_u}^2 ,\
\e_{L_2}^2\e_{H_u}^2 ,\
\e_{L_1}^2 \e_{H_u}^2 \right), \eea as well as the
mixing angles \beq |V_{\mathrm{CKM}}| \approx \left(
\begin{array}{ccc}
1 & \e_{Q_1}/\e_{Q_2} & \e_{Q_1}/\e_{Q_3} \\
\e_{Q_1}/\e_{Q_2} & 1 & \e_{Q_2}/\e_{Q_3} \\
\e_{Q_1}/\e_{Q_3} & \e_{Q_2}/\e_{Q_3} & 1 \\
\end{array}
\right),\,\,\,
|V_{\mathrm{MNS}}| \approx \left(
\begin{array}{ccc}
1 & \e_{L_1}/\e_{L_2} & \e_{L_1}/\e_{L_3} \\
\e_{L_1}/\e_{L_2} & 1 & \e_{L_2}/\e_{L_3} \\
\e_{L_1}/\e_{L_3} & \e_{L_2}/\e_{L_3} & 1 \\
\end{array}
\right).
\eeq
This generic structure can well accommodate the present data.  For
example, the observed mixing angles (taken from \cite{pdg} and
\cite{GonzalezGarcia:2009ij}) are given by
\beq\label{eq:CKM}
|V_{\mathrm{CKM}}| \simeq \left(
\begin{array}{ccc}
0.97 & 0.23 & 0.004 \\
0.23 & 0.97 & 0.04 \\
0.009 & 0.04 & 0.99 \\
\end{array}
\right),\,\,\,
|V_{\mathrm{MNS}}| \simeq \left(
\begin{array}{ccc}
0.77\,\mbox{--}\,0.86 & 0.50\,\mbox{--}\,0.63 & 0.00\,\mbox{--}\,0.22 \\
0.22\,\mbox{--}\,0.56 & 0.44\,\mbox{--}\,0.73 & 0.57\,\mbox{--}\,0.80 \\
0.21\,\mbox{--}\,0.55 & 0.40\,\mbox{--}\,0.71 & 0.59\,\mbox{--}\,0.82 \\
\end{array}
\right).
\eeq
It is particularly interesting to note that this generic structure
predicts that these matrices are symmetric up to $O(1)$ effects, and
hence prefers a larger value of $\theta_{13}$ in the neutrino sector.

In this paper, we'll focus primarily on supersymmetric $\SU(5)$
GUT models, with three SM generations $T_i\in \tenrep$ and $\bar
F_i\in\fivebarrep$, and Higgs fields $H\in\fiverep, \bar H \in
\fivebarrep$.  When $\SU(5)_{\GUT}$ relations are satisfied at the
unification scale $M_{\GUT}\sim 10^{16} \gev$, one has the
relations
\beq
\epsilon_{Q_i} = \epsilon_{U_i} = \epsilon_{E_i}
\equiv \epsilon_{T_i}
\quad\mbox{and}\quad
\epsilon_{D_i} = \epsilon_{L_i} \equiv \epsilon_{\bar{F}_i}.
\eeq
Considering the GUT values of the up-type quark masses then yields the
estimates (extracted from Ref.~\cite{Antusch:2008tf} at $\tan\beta \sim 30$)
\beq \epsilon_{T_i}
\sqrt{\epsilon_{H}} \approx \left(.001\,\mbox{--}\,.002, .03\,\mbox{--}\,.04, .7\,\mbox{--}\,.8\right). \eeq
Note however that a somewhat larger value of $\epsilon_{T_1}$ (by a factor of $\sim 4\,\mbox{--}\,5$)
is preferred by the Cabibbo angle in \Eq{eq:CKM}.

Similarly, considering the GUT values of the down-type quark
masses yields the estimates
\beq\label{eps5quarks}
\epsilon_{\bar{F}_i} \epsilon_{\bar{H}} \approx
\tan\beta\times\left(.002\,\mbox{--}\,.01, .001\,\mbox{--}\,.007, .006\,\mbox{--}\,.02\right), \eeq
while considering the lepton masses gives \beq\label{eps5leptons}
\epsilon_{\bar{F}_i} \epsilon_{\bar{H}} \approx
\tan\beta\times\left(.001\,\mbox{--}\,.002, .006\,\mbox{--}\,.01,
.01\,\mbox{--}\,.03\right). \eeq  Here we have tried to illustrate the
potentially large uncertainties arising from SUSY threshold corrections,
which can be important at larger values of $\tan\beta$.  In most cases, an $O(1)$ violation of
the GUT relations in the first and second generations will be required to be consistent with
low-energy data.  Nevertheless, it is interesting that the estimates
of \Eqs{eps5quarks} and (\ref{eps5leptons}) are not hierarchically
different from one another.  In particular, it is still entirely
possible that the suppression factors arising from the K\"{a}hler
potential respect $\SU(5)_\GUT$ relations but the anarchical
superpotential couplings do not.  Moreover, it may still be possible that
$\SU(5)_\GUT$ relations are allowed in the presence of large SUSY threshold
corrections~\cite{DiazCruz:2000mn,Ibrahim:2003ca,Ferrandis:2004ri,DiazCruz:2005ri,Ross:2007az,Maekawa:2007gr,Antusch:2008tf,Enkhbat:2009jt}.
In the present work, we will not present a completely realistic GUT model or
discuss the origin of SUSY-breaking parameters, and so will for the most part
attempt to remain agnostic about how $O(1)$ violations of these relations may arise.

Perhaps the minimal potentially realistic structure consistent with
these estimates is a `{\bf 10}-centered' model, or a model which
generates suppressions in $\epsilon_{T_1}$ and $\epsilon_{T_2}$.  In
the simplest version, the smallness of the down-type masses must be
generated through a combination of large $\tan\beta$ and somewhat
small $O(1)$ factors.  Smaller $\tan\beta$ is possible, however, if
the model additionally generates suppressed values of
$\epsilon_{\bar{H}}$ or $\epsilon_{\bar{F}_i}$.  In this case, a
somewhat small value of $\epsilon_{\bar{F}_1}$ relative to
$\epsilon_{\bar{F}_{2,3}}$ may also be preferred to explain the
smallness of $m_e/m_{\mu}$.  However, since this may not be necessary
to explain the ratio of $m_d/m_s$, this will depend on the details of
how GUT-breaking effects are introduced.

While somewhat tangential to the present work, it is interesting to
note that the above approach to flavor physics can have important implications
for the SUSY flavor problem.  If the flavor hierarchies simply arise as
small numbers in the superpotential (as is commonly assumed), one might worry that generic $M_\pl$-suppressed
SUSY-breaking operators in the K\"{a}hler potential such as $X^{\dagger} X \Phi_i^{\dagger} \Phi_j /
M_\pl^2$ will lead to large flavor violation in the soft masses if they
are the dominant contribution.  In this case one is strongly motivated to
think about alternative mediation mechanisms such as gauge
mediation~\cite{Dine:1981gu,Dine:1994vc}.  On the other hand, if the
flavor hierarchies arise from the wave-function factors $Z_i$, then
supersymmetry-breaking operators will also be suppressed by factors of
$\epsilon_i$ upon canonically normalizing the fields, potentially
leading to a simple solution of the flavor
problem~\cite{Nelson:2000sn,Nelson:2001mq,Kobayashi:2001kz,Kitano:2006ws,Nomura:2007ap,Nomura:2008gg}.
Note that in the present scenario these operators can also be further
suppressed (or enhanced) because in superconformal theories the
anomalous dimension of $\Phi^{\dagger}_i \Phi_j$ need not be simply
related to the anomalous dimension of $\Phi_i$.

\subsection{Superconformal Flavor Models}
\label{sec:scfts}

The models we will consider in this paper are based on an elegant
solution for producing hierarchical suppression factors using strong
dynamics, which was first proposed in a supersymmetric context by
Nelson and Strassler in~\cite{Nelson:2000sn}.  They considered extending
the MSSM to include a strongly coupled sector with gauge group $G$.
Standard Model fields $\Phi_i$ are singlets under $G$, but can develop
large anomalous dimensions $\g_i$ via superpotential couplings to
operators in the new sector.

These anomalous dimensions exponentiate into large wave-function
factors in the IR, leading to suppression factors
\beq
\label{eq:wavefunctionrunning}
\frac{d \log Z_i}{d\log \mu}=-\g_i
\qquad\Longrightarrow\qquad
\e_{\Phi_i}= \exp\p{-\half \int_{\log \L_c}^{\log\L} \g_i\, d\log\mu},
\eeq
where $\L$ is the UV scale at which the Yukawas are anarchical, and
$\L_c$ is the scale at which the exotic sector decouples.  The
anomalous dimensions $\g_i$ are typically $O(1)$.  So to
satisfactorily reproduce the SM Yukawa hierarchies we need the strong
dynamics to persist over a large range of scales $\L_c<\mu<\L$.
Following \cite{Nelson:2000sn}, we consider the case where the strong
sector gauge and superpotential couplings flow from the UV to an
approximate conformal fixed point at $\mu = \L$ and exit the
approximately conformal regime at $\mu=\L_c$, for instance due to
$\L_c$-scale masses for the exotic matter charged under $G$.

In the range $\mu\in[\L_c,\L]$, we can compute the anomalous
dimensions $\g_i$ at leading order by ignoring the SM gauge and Yukawa
couplings and approximating our theory as a supersymmetric conformal
field theory (SCFT) with strongly-coupled gauge group $G$ and an $\SU(5)$ global flavor
symmetry.  In this approximation, the $\g_i$ are constant, and
\Eq{eq:wavefunctionrunning} can be integrated trivially to give
\beq
\label{eq:suppressionfactors}
\e_{\Phi_i} = \p{\frac{\L_c}{\L}}^{\half\g_{i}}\quad\mbox{for each $\Phi_i=T_i,\bar F_i, H,\bar H$.}
\eeq

Conformal or near-conformal dynamics is often easiest to understand in
a `physical' basis where we keep the fields canonically normalized
$\sqrt{Z_i}\Phi_i\to \Phi_i$ at each scale $\mu$, rather than a
`holomorphic' basis where the superpotential is
not renormalized.\footnote{For instance, in a holomorphic basis, gauge
  couplings run at one loop, while kinetic terms run according to
  \Eq{eq:wavefunctionrunning}.  At a conformal fixed point, both of
  these effects are nonzero, but the rescaling anomaly~\cite{ArkaniHamed:1997mj} from passing to
  the physical basis exactly cancels the one-loop gauge-coupling
  running.  It's simpler to stay in the physical basis where neither
  the kinetic terms nor the gauge couplings run at a conformal
  fixed point.}
We will use the physical basis for the rest of this paper, so let us
take a moment to recover \Eq{eq:suppressionfactors} using it.  In the
physical basis, superpotential couplings run according to the
anomalous dimensions of the associated operators, while kinetic terms
in the K\"ahler potential are RG invariant.  For instance, the
Standard Model Yukawa couplings $y_u^{i j}$ and $y_d^{ij}$ run
according to
\bea
\label{eq:yukawarunning}
\frac{d y^{ij}_u}{d\log\mu} &=& \b_{y^{ij}_u} \,=\, \half(\g_{T_i} + \g_{T_j} + \g_{H}) y^{ij}_u ,\nonumber\\
\frac{d y^{ij}_d}{d\log\mu} &=& \b_{y^{ij}_d} \,=\, \half(\g_{ T_i} + \g_{\bar F_i} + \g_{\bar H}) y^{ij}_d.
\eea
In the CFT approximation where $\g_i$ are constant, these integrate to
\bea
y^{ij}_u(\L_c) &=& \p{\frac {\L_c}{\L}}^{\half{(\g_{T_i} + \g_{T_j} + \g_{H})}} y^{ij}_u(\L), \nonumber\\
y^{ij}_d(\L_c) &=& \p{\frac {\L_c}{\L}}^{\half{(\g_{T_i} + \g_{\bar F_j} + \g_{\bar H})}} y^{ij}_d(\L),
\eea
which of course agrees with the suppression factors
\Eq{eq:suppressionfactors}.

\subsubsection{Toy Model}
A particularly attractive feature of superconformal flavor models is that different SM
generations can be singled out dynamically by the CFT.  Indeed, even
with flavor anarchy in the UV, the basic representation theory of the
exotic sector can naturally lead to a hierarchy.  As an example, let
us consider a toy strong sector with $G=\SU(N)$ and matter content
given in Table~\ref{tab:onegeneration}.

\begin{table}[h]
\begin{center}
\begin{tabular}{ | l | c | c | }
\hline
                    & $\SU(5)_\GUT$ & $\SU(N)$ \\
\hline
$X+S$          & $\tenrep + \trivrep$           & $\Yfund$ \\
$\bar X+\bar S$    & $\tenbarrep + \trivrep$      & $\Yafund$ \\
\hline
\end{tabular}
\end{center}
\caption{Matter content of a toy model with one light generation.}
\label{tab:onegeneration}
\end{table}

Before coupling to the MSSM (and treating $\SU(5)_\GUT$ as a
spectator), this is simply $\SU(N)$ supersymmetric QCD with $N_f=11$,
which flows to an interacting conformal fixed point when $\frac 3 2 N
< 11 < 3 N$~\cite{Seiberg:1994pq}, or equivalently $4\leq N\leq 7$.
However, in addition to the usual SUSY QCD fixed point, we we can
reach other non-trivial fixed points by deforming the theory with
relevant operators.  For example, we can add a relevant operator $h(\mu)\mu^{3-d}\cO$ to the
superpotential at a UV scale $\mu$, where $d$ is the classical dimension of $\cO$ and we have
normalized $h(\mu)$ to be dimensionless.  One generally then expects the (initially small) coupling
$h(\mu)$ to grow towards the IR, until it approaches a zero of
\beq
\b_h &=& (\dim \cO - 3)\,h\ \ =\ \ \p{d+\frac{\g_\cO}{2}-3}h.
\eeq
That is, $h(\mu)$ will grow (and other couplings might also change) until the operator $\mu^{3-d}\cO$ becomes
approximately marginal ($\dim \cO \simeq 3$) and we've reached the vicinity of a new conformal fixed point,
with $h(\mu) \simeq h_*$.

Our toy exotic sector admits exactly one relevant gauge-invariant
coupling to the MSSM,
\beq
W_{int} &=& h\, T_1 \bar X S.
\eeq
Note that while there may be flavor anarchy in the UV, only one linear
combination of the $T_i$ can participate in a relevant coupling to the
exotic sector, and we have simply {\it defined} that linear
combination to be $T_1$ without loss of generality.  Now suppose our
theory flows to a fixed point where the operator $T_1\bar X S$ becomes
marginal.  In the next section, we will show that $T_1$ has a positive
anomalous dimension at this fixed point and thus develops a
suppression factor through RG running.  The other two generations
$T_2,T_3$ don't couple to the exotic sector, so they have vanishing
anomalous dimensions in the SCFT approximation and don't develop
suppression factors.  Thus this toy model has exactly one light
generation.

More formally, the flavor structure is as follows: given an $\SU(3)_F$
flavor symmetry among the $T_i$ in the UV (broken only by $O(1)$
Yukawas), we can think of the coupling $h$ as a spurion in the
$\bar{\bf 3}$ of $\SU(3)_F$.  Picking a direction for $h$ breaks
$\SU(3)_F\to \SU(2)$, singling out a generation.  This breaking is
$O(1)$ at the scale $\L$, but gets exponentiated by
conformal running.  This is analogous to warped flavor models, where
bulk masses provide an $O(1)$ source of flavor breaking in addition to
the Yukawas, but they get exponentiated in the AdS wave-function
profiles.  However, as noted in the introduction, bulk masses are
typically {\it inputs} of warped flavor models, whereas anomalous
dimensions (and the resulting Yukawa hierarchies) are {\it outputs} of
the models we consider here.

To make quantitative predictions for flavor physics, we'd like to
actually determine the anomalous dimensions of the SM fields that
couple to the strong sector.  When the scenario of
\cite{Nelson:2000sn} was first proposed, the problem of finding
scaling dimensions of chiral operators in an SCFT was unsolved in
general, and a number of candidate superconformal flavor models
appeared to be `incalculable'.  However since then, Intriligator and
Wecht~\cite{Intriligator:2003jj} found a simple and widely applicable
solution called $a$-maximization, which we'll now review.

\subsection{Review of $a$-Maximization}
\label{sec:amax}

The $\cN = 1$ superconformal symmetry group in 4 dimensions is
$\SU(2,2|1)$, which has $\SO(2,4)\x \U(1)_R$ as its bosonic subgroup.
Thus, every SCFT necessarily has a distinguished non-anomalous
$\U(1)_R$ charge related by supersymmetry to other generators of the
conformal group.  In particular, the scaling dimension of a chiral
operator $\cO$ is determined by its $R$-charge, $\dim(\cO)=\frac 3 2
R(\cO)$.

Suppose a SCFT has a $\U(1)$ symmetry $R_0$ under which the SUSY
generators $Q_\a$ have charge $-1$, and a collection of $\U(1)$
``flavor" symmetries $F_I$ under which the $Q_\a$ are neutral. Any
non-anomalous linear combination \beq R_t = R_0 + \sum_I s_I F_I \eeq
is a candidate $R$-symmetry in the sense that the SUSY generators have
the appropriate charge under $R_t$ (`$t$' stands for `trial').
However, only one is the distinguished $R$-symmetry appearing in the
superconformal group.  In simple cases, we can identify the correct
linear combination from symmetry considerations alone.  The
prototypical example is supersymmetric QCD with $\frac 3 2 N_c < N_f <
3 N_c$ and zero superpotential, which flows to a conformal fixed point
in the IR, and has only a single non-anomalous $R=R_t$ that commutes
with all flavor symmetries.  In the UV, there is nothing special about
this $R$, but in the deep IR, it becomes part of the emergent
$\SU(2,2|1)$ superconformal symmetry and determines the scaling
dimensions of chiral operators.

Theories with less symmetry generally have a non-trivial affine space
of possible $R_t$'s.  However, a simple condition
\cite{Intriligator:2003jj} uniquely determines the correct
superconformal $R$-charge: it is a local maximum of
\beq
\label{eq:amaximization} a(R_t) &=&
\frac{3}{32}\left[3\Tr(R_t^3)-\Tr(R_t)\right],
\eeq
where $\Tr(R_t^3)$ and $\Tr(R_t)$ are the coefficients of the gauge
anomaly $\<\ptl_\mu J^\mu_{R_t}\, J^\nu_{R_t} J^\rho_{R_t}\>$ and
gravitational anomaly $\<\ptl_\mu
J^\mu_{R_t}\,T_{\nu\rho}\,T_{\sigma\gamma}\>$, respectively.  This
condition is called `$a$-maximization.'  We review the proof in
Appendix \ref{sec:proofofamax}.

When our CFT arises as the fixed point of a theory that is weakly
coupled in the UV, one can often reliably compute the traces in
perturbation theory in the weakly-coupled description using the 't
Hooft anomaly-matching conditions.  An important exception to this
occurs when accidental $\U(1)$ symmetries emerge in the IR which
aren't manifest in the UV Lagrangian\,---\,in this case one must also include
these accidental symmetries when maximizing $a(R_t)$.

One possible signal that accidental symmetries are arising is that a
gauge-invariant operator $\mathcal{O}$ in the chiral ring appears to
violate the bound required by unitarity, given by $R_{\mathcal{O}}
\geq 2/3$ for scalar operators~\cite{Mack:1975je}.  In this case a
reasonable interpretation is that the operator $\cO$ is becoming a
free field (with $R_{\mathcal{O}} = 2/3$)~\cite{Seiberg:1994pq}
 and that an accidental
symmetry associated with rotations of $\cO$ is emerging in the IR.  A
useful description of this phenomenon~\cite{Barnes:2004jj} that makes
the emergent symmetry manifest involves introducing an additional
vector-like pair of fields $\{L,M\}$ with superpotential
\beq
W_{LM} &=& L(M-\cO).
\eeq
When $\cO$ is consistent with the unitarity bound, we can think of
this as a (somewhat trivial) dual description of the theory we started
with.  $L$ and $M$ are massive, so we could just integrate
them out out.  Equivalently, when the operator $LM$ becomes marginal,
we have $R(L)=2-R(M)$, which means that the contributions of $L$ and
$M$ cancel in the $a$-maximization calculation, so they don't
influence the $R$-charges of other fields.

The theory with $L$ and $M$ is presumably identical to the original in
the deep IR.  But when $\cO$ violates the unitarity bound, we can now
describe what's going on.  The coupling $LM$ flows to zero, leaving us
with a free field $M$ with $R$-charge $2/3$.  Meanwhile, the nonzero
coupling $L\cO$ sets $\cO$ to zero in the chiral ring, resolving any
conflict with the unitarity bound.  The field $L$ has $R(L)=2-R(\cO)$,
and its contribution to $a$ no longer cancels with $M$.  Including $L$
and $M$ in the $a$-maximization procedure requires the simple
modification~\cite{Anselmi:1997ys,Kutasov:2003iy}
\bea\label{eq:amodification}
a(R_t) &\rightarrow& a(R_t) + a(M) + a(L) \nonumber\\
                 &=& a(R_t) + a(2/3) - a(\cO) \nonumber\\
                 &=& a(R_t) + \frac{\dim(r_{\cO})}{96} (2 - 3 R_{\cO})^2 (5 - 3 R_{\cO}),
\eea
where $r_{\cO}$ denotes the representation of $\cO$.  It's important to
note that $L$ and $M$ can contribute to other anomalies as well.
For instance, if $\cO$ is in a non-trivial representation of
$\SU(5)_\GUT$, we must include $L$ and $M$ in calculating $\b_{g_5}$
whenever $\cO$ violates the unitarity bound.

Describing this special case of a gauge-singlet operator becoming free
is relatively simple.  More generally, however, it is not always easy
to determine when accidental symmetries arise, and it is important to
study carefully any known dual descriptions of the theory under
consideration in order to gain evidence for their emergence.  See, e.g.,~\cite{Kutasov:2003iy,Intriligator:2003mi,Kutasov:2003ux,
Barnes:2004jj,Barnes:2005zn,Csaki:2004uj,Intriligator:2005if,Kawano:2005nc,Kawano:2007rz,Shapere:2008un,Poland:2009px}
for additional discussion and many examples.

\subsubsection{$a$-Maximization in the Toy Model}

As a simple example of $a$-maximization, let us consider the toy model
in Table~\ref{tab:onegeneration}, with $R_t$-charges $R_X,R_{\bar X},
R_S, R_{\bar S},R_{T_1}$ for the superfields.  ($T_2, T_3$, and
the rest of the MSSM do not couple to the strong sector, so they are
free fields with $R = 2/3$.)  In the UV, gauge and gravitational
anomalies come from the gauginos which have $R_t$-charge $1$, and from
the matter fermions which have $R_t$-charges 1 less than their
associated chiral superfields.  Thus,
\beq \label{eq:aexample} a(R_t)&=&
\frac{3}{32} \left[ 2(N^2-1)+\sum_{i}
\dim(r_i)\left(3(R_i-1)^3-(R_i-1)\right) \right],
\eeq where the index $i$ runs over all of the chiral superfields, and
$r_i$ is the representation of each field.

The trial $\U(1)_R$ should be anomaly-free with respect to $G$, so
we have the constraint \beq \label{eq:nonanomalousexample}
0 &=& T(G)+\sum_i (R_i-1)T(r_i) \nonumber\\
&=& N + 5(R_X - 1) + 5(R_{\bar{X}} -1) + \frac12(R_S - 1) +
\frac12(R_{\bar{S}}-1), \eeq where $T(r)$ is the Dynkin index of
$r$ as a $G$-representation.  Using $\frac 3 2 R_i=1+\frac{\g_i}{2}$, we see that
(\ref{eq:nonanomalousexample}) is equivalent to vanishing of the
numerator of the exact NSVZ $\b$-function~\cite{NSVZ:1983,SV:1986}
\beq\label{eq:NSVZ} \b_{g_G} &=&
\frac{-\left(3T(G)-\sum_i(1-\g_i)T(r_i)\right)}{16\pi^2\left(1-\frac{T(G)g_G^2}{8\pi^2}\right)}g_G^3.
\eeq
That is, $R$ is conserved if and only if the theory is conformal,
which reflects the fact that supersymmetry relates the $R$-current
anomaly $\ptl_\mu J_R^\mu$ to the trace of the
stress-energy tensor $T_\mu^\mu$.

A second constraint on $R_t$ comes from the fact that the coupling
$T_1\bar X S$ is exactly marginal in our CFT.  Thus, we have
\beq
2 &=& R_{T_1}+R_{\bar X}+R_S.
\eeq
Finally, numerically minimizing (\ref{eq:aexample}) subject to our two
constraints yields the $R$-charges in
Table~\ref{tab:toyRcharges}.  Note that $R_{T_1}$ increases with $N$,
reflecting the fact that supersymmetric QCD becomes more strongly
coupled as $N\to 2N_f/3$, pulling the anomalous dimensions of the
exotic quarks more negative.  We could alternatively analyze our
theory in a Seiberg dual description where the coupling $T_1 \bar X S$
becomes a mass coupling to a meson $T_1 M_{\bar X S}$.  The magnetic
description becomes more weakly-coupled as $N\to 2N_f/3$,
and $R_{M_{\bar X S}}$ approaches its free value $2/3$.  The
$a$-maximization procedure is identical in this dual description,
since the two theories have matching anomalies.  We present
further discussion of the dual description in Appendix~\ref{sec:CFTexit}.

\begin{table}[t]
\begin{center}
\begin{tabular}{ c | c | c | c | c | c}
$N$ & $R_{T_1}$ & $R_X$ & $R_{\bar X}$ & $R_S$ & $R_{\bar S}$\\
\hline
$4$ & $.686$ & $.632$ & $.637$ & $.677$ & $.632$ \\
$5$ & $.771$ & $.683$ & $.546$ & $.533$ & $.533$ \\
$6$ & $.920$ & $.625$ & $.455$ & $.439$ & $.439$ \\
$7$ & $1.191$  & $.445$ & $.364$ & $.356$ & $.356$ \\
\end{tabular}
\end{center}
\caption{$R$-charges of our toy model with one light generation (Table~\ref{tab:onegeneration}).}
\label{tab:toyRcharges}
\end{table}

\section{Vector-like Models}
\label{sec:sunmodels}

In this section we will attempt to find the simplest superconformal
models that display a realistic flavor structure.  For simplicity we
will primarily consider conformal sectors that are vector-like.  This
can easily ensure that the conformal sector exits at a high scale and
that all exotic states decouple, because one can simply write down
(large) mass terms for all of the fields.\footnote{By contrast,
  \cite{Nelson:2000sn} focused on chiral strong sectors, with the
  CFT-breaking scale $\L_c$ generated dynamically by some additional
  gauge group.  This is a promising way to avoid a hierarchy problem
  for $\L/\L_c$, but the models are more complicated and there is no
  general guarantee that exotic states will be massive.}  For example,
in the toy model discussed in the previous section we can write down
the mass terms $W \supset m_{X} X \bar{X} + m_S S \bar{S}$.  Assuming
that $m_X$ becomes marginal first, we expect the CFT regime to exit at
a scale $\mu \sim \L_c$, where $\L_c = m_X (\L_c / \L)^{\frac12
  (\g_X + \g_{\bar{X}})}$.  Meanwhile, below the scale $\mu \sim m_S
(\L_c / \L)^{\frac12 (\g_S + \g_{\bar{S}})}$ all exotic states will
have decoupled and we are left with the MSSM at low energies.

Furthermore, we will mainly consider theories with an $\SU(5)_\GUT$
symmetry given their many successful predictions.  We will attempt to
remain agnostic about the details of GUT-breaking and doublet-triplet
splitting, as these issues are (for the most part) decoupled.  As
discussed in Section \ref{sec:flavor}, the minimal quasi-realistic
structure in the context of $\SU(5)_\GUT$ is that of a `{\bf
  10}-centered' flavor model, or a model that generates different
suppression factors for two matter fields in the
$\tenrep$-representation.  We will start by finding models in this
class.  In several cases, the models can be extended easily to
generate suppression factors for one or more generations of
$\fivebarrep$'s, or possibly the $\fivebarrep$ Higgs.

As a minimum requirement, the conformal sector should then contain two
singlet operators $\mathcal{O}_{1,2}$ in the $\tenbarrep$
represenation of $\SU(5)_\GUT$, so that one can introduce couplings
\beq\label{eq:wintgeneric} W_{int} = T_1 \mathcal{O}_1 + T_2
\mathcal{O}_2. \eeq These interactions should be marginal
couplings of the CFT in order for $T_{1,2}$ to acquire large anomalous
dimensions, and $\mathcal{O}_1$ and $\mathcal{O}_2$ should have
different dimensions under the CFT in order to explain the hierarchy.
In particular, there should be no symmetry relating them.  On the
other hand, we would also like to have evidence that the CFT exists
and that there is a reasonable flow that could lead to the desired
fixed point.  To this end, we would like these operators to be
relevant deformations of the CFT in which they are absent.  Typically
this requires that the coupling $T_i\cO_i$ involve at most three or
four fields.  Of course, CFT gauge interactions can lower the
dimensions of strong-sector fields, but in many models one will leave
the conformal regime or run into violations of the unitarity bound
before the 5-field operators can become relevant.

At what scale should the CFT exit?  One strong constraint arises if
the interactions in \Eq{eq:wintgeneric} violate baryon and lepton
number.  This will be the case in the simplest models that we will
consider.  In this situation integrating out the CFT states around the
scale $\L_c$ will then induce dimension 6 proton-decay operators in
the K\"ahler potential suppressed by $\sim 16\pi^2/(N\L_c^2)$.  This implies that
$\L_c$ should be near or above the GUT scale in these
models.\footnote{On the other hand, it is interesting to note that
  superconformal flavor models can actually improve the situation with
  dimension 5 proton-decay operators, because they will also receive
  suppression from the $\epsilon_i$ factors.  We refer readers
  to~\cite{Nelson:2000sn} for a more thorough discussion of these
  operators.}  There is then a rather small allowed window for
conformal running, \beq M_\GUT\lsim \L_c<\L\lsim M_\pl,\eeq which in
turn means that the theory will need to be fairly strongly-coupled in
order to reproduce the observed hierarchies.

Since the CFT sector will introduce many new fields in
representations of $SU(5)_\GUT$, an additional strong constraint
on these models comes from requiring that the GUT gauge
coupling remain perturbative throughout the conformal window.
Since some of these fields will have large anomalous dimensions,
we should integrate the full NSVZ $\beta$-function
\Eq{eq:NSVZ}, which we will parameterize as \bea
\label{eq:NSVZrewrite} \beta_{g_5} =
\frac{\mathcal A}{16\pi^2 \left(1
- \frac{5g_{5}^2}{8\pi^2}\right)} g_5^3, \eea
where
\beq
\mathcal A \equiv
-3\,\Tr\left[\U(1)_R\,\SU(5)_\GUT^2 \right] = - 15 + \sum_i (1 - \g_i) T(r_i).
\eeq
It is clear that we can only determine the evolution of $g_5$ after we
know the correct $\U(1)_R$ symmetry, which can in turn only be
determined by performing $a$-maximization.  However, if we assume that
the CFT exits and all exotic states charged under $\SU(5)_\GUT$
decouple near $\L_c \sim 10^{16}~\gev$, we can still place a
model-independent bound on $\L$ as a function of $\mathcal A$.

\begin{figure}
\begin{center}
\includegraphics[scale=1.4]{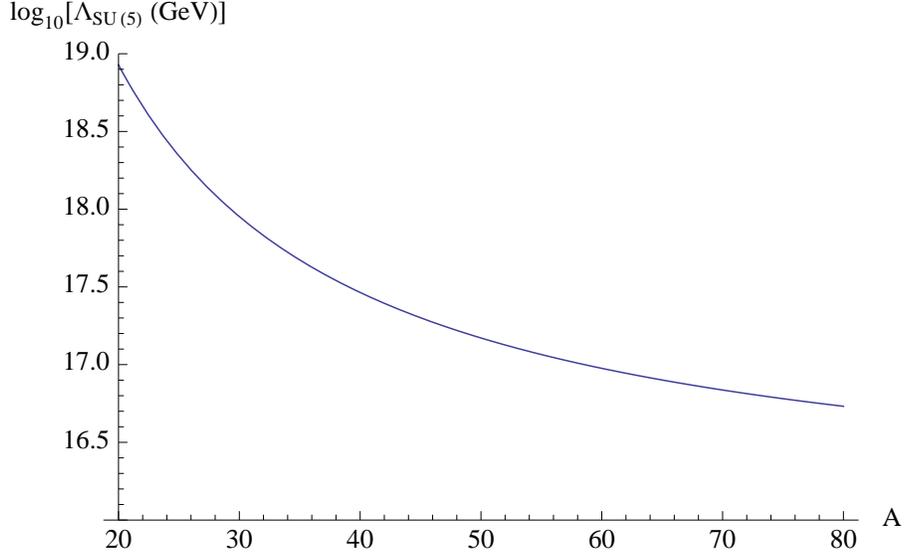}
\end{center}
\caption{Location of the $SU(5)_{\GUT}$ Landau pole as a function
of the $\beta_{g_5}$-numerator $\mathcal A$, assuming that all new
matter enters at $\L_c \sim 10^{16} \gev$.}
\label{su5bound}
\end{figure}

Integrating \Eq{eq:NSVZrewrite} gives a Landau pole $\L_{\SU(5)}$ at
\beq
\label{eq:landaupole}
\log\frac{\L_{\SU(5)}}{\L_c}
&=&
\frac{5}{\mathcal A}\p{\frac{8\pi^2}{5g_5^2(\L_c)}+\log\frac{5g_5^2(\L_c)}{8\pi^2}-1}.
\eeq
The suppression factors $\e_i$ should be generated well below
$\L_{\SU(5)}$, since otherwise the approximation of our theory as a
weakly perturbed SCFT breaks down.  If we had an appropriate dual
description of the $\SU(5)_\GUT$ group, it might be possible to make
sense of a scenario where conformal running persists through the
Landau pole.  But we would likely still lose any semblance of a
simple, predictive solution to the flavor problem.  Thus, we will
demand that the conformal running distance $\log\frac{\L}{\L_c}$ not
exceed that of \Eq{eq:landaupole}.  We plot this bound in
Figure~\ref{su5bound}, where we have taken $\alpha_5(\L_c) = g_5^2(\L_c)/{4\pi} = 1/25$ as
determined from running up the low-energy gauge couplings.  The curve
indicates the location of $\L_{\SU(5)}$.  We will find that this
constraint rules out a number of otherwise viable models, because they
would require $\L$ to be larger than allowed by this bound in order to
reproduce the observed hierarchies.

Below we will attempt to construct the simplest realistic models
that avoid the above constraints.  We will start by considering
the simplest extension of the toy model that can accommodate two
light generations, but we will find that the constraint from the
$\SU(5)_{\GUT}$ Landau pole is too strong.  This will motivate
looking for models containing only $\fiverep$ and $\fivebarrep$
representations of $\SU(5)_{\GUT}$ in order to minimize the
contribution to $\beta_{g_5}$ from the CFT sector.  We will then
analyze the simplest such models based on $\SU(N)$, $\SO(N)$, and
$\Sp(2N)$ gauge groups.

\subsection{\texorpdfstring{$\tenrep+\fivebarrep+\trivrep$ Model}{10+5bar+1 Model}}

We will start by extending our toy model from Section \ref{sec:scfts}
to generate suppression factors for two generations.  An obvious way
to get two couplings to $T_i$'s is to add a second pair of GUT
singlets $S'+\bar S'\in (\trivrep, \Yfund)\oplus (\trivrep,\Yafund)$,
giving us another $\tenrep\,\tenbarrep\,\trivrep$ coupling: $T_2 \bar
X S'$.  However, this will not work because there is a symmetry
relating $S$ and $S'$ that would ensure that $T_1$ and $T_2$ have the
same anomalous dimension.  The only other types of 3-field GUT
couplings to $\tenrep$'s are are $\tenrep\,\tenrep\,\fiverep$ and
$\tenrep\,\fivebarrep\,\fivebarrep$.  We can get the first of these by
adding a $(\fiverep, \Yafund)\oplus(\fivebarrep, \Yfund)$ pair, leaving us
with the matter content of Table~\ref{tab:tenfiveone}.

\begin{table}[h]
\begin{center}
\begin{tabular}{ | l | c | c | }
\hline
                    & $\SU(5)_\GUT$ & $\SU(N)$ \\
\hline
$X+\bar{Q}+S$       & $\tenrep + \fivebarrep + \trivrep$ & $\Yfund$ \\
$\bar{X}+Q+\bar{S}$ & $\tenbarrep + \fiverep + \trivrep$ & $\Yafund$ \\
\hline
\end{tabular}
\end{center}
\caption{Matter content of the $\tenrep + \fivebarrep + \trivrep$
model.} \label{tab:tenfiveone}
\end{table}

Under the $\SU(N)$ gauge theory, there are $N_f = 16$ flavors of
vector-like quarks.  Thus, the theory with vanishing
superpotential is in an interacting conformal regime for $3N/2 <
16 < 3 N$, or equivalently $6\leq N\leq 10$.  Meanwhile, from the
point of view of $\SU(5)_\GUT$ we have added $N$ vector-like pairs
of generations, and we should be concerned about the constraint
from the $\SU(5)_{\GUT}$ Landau pole.  Indeed we'll find shortly
that this model cannot account for the observed hierarchy if $g_5$
remains perturbative in the conformal window.

In addition to the interactions $T_i X Q$ and $T_i \bar X S$, this
sector admits other couplings to the Standard Model.  The ones
that are relevant at weak coupling are the 3-field
operators $\bar{F}_i Q S$, $\bar{H} Q S$, and $H \bar{Q} \bar{S}$.
In addition, a number of 4-field operators can potentially become
relevant at strong coupling.  However, in order to forbid dangerous
dimension $3$ and $4$ lepton- and baryon-number violating
operators, a realistic theory requires an additional approximate
(discrete or continuous) symmetry, which may in turn forbid some
subset of the allowed deformations.  Here we will simply focus on
the interactions that are the most interesting for flavor
physics, giving possible symmetries (where appropriate) that would
forbid the remaining operators.

The minimal superpotential we need for a $\tenrep$-centered model
is \beq\label{eq:sup-t1t2-1051} W_{\tenrep} = T_1 X Q + T_2 \bar{X} S.
\eeq  As before, without loss of generality we can simply define
the linear combinations of matter fields that couple to the above
CFT operators to be $T_1$ and $T_2$.  If the theory has, e.g., an
approximate $\U(1)_R$ symmetry in the UV under which $\left\{H, \bar{H}, Q,
S\right\}$ have charge $0$, $\left\{ T_i, \bar{F}_i, N_i, X,
\bar{X}\right\}$ have charge $1$, and
$\left\{\bar{Q},\bar{S}\right\}$ have charge 2, then these are the
only allowed interactions.\footnote{Note that this is similar to
the symmetry proposed in~\cite{Hall:2002ci}, which was in part
motivated by a possible solution to the doublet-triplet splitting
problem.  This symmetry is not to be confused with the superconformal
$\U(1)_R$ that becomes important during conformal running.  Rather,
here we are imagining that the theory above the conformal regime has an
approximate symmetry that causes some operators to have smaller
coefficients than others.}

Now we will determine the $\U(1)_R$ symmetry using $a$-maximization.
The superpotential \Eq{eq:sup-t1t2-1051} and anomaly
cancelation impose the constraints \bea\label{eq:constenfiveone}
2 &=& R_{T_1} + R_X + R_Q \nonumber\\
2 &=& R_{T_2} + R_{\bar{X}} + R_S \\
0 &=& N + 5 (R_X - 1) + 5 (R_{\bar{X}}-1) + \frac52 (R_Q - 1) +
\frac52 (R_{\bar{Q}}-1) + \frac12 (R_S-1) + \frac12
(R_{\bar{S}}-1). \nonumber\eea Maximizing $a(R_t)$ subject to
these constraints then yields the $R$-charges given in
Table~\ref{tab:rchargetenfiveone}.

In the table (and throughout the paper), we have defined the
phenomenologically required conformal window $\L/\L_c$ to be the
running distance required to generate suppression factors that are
simultaneously within a factor of 3 of the values $\e_{T_1} = .003$
and $\e_{T_2} = .04$.  The window on $\e_{T_1}$ covers both the somewhat
smaller suppression factor preferred by the up quark mass, as well as the
somewhat larger suppression factor preferred by the Cabibbo angle.
In addition, we've indicated the numerator of $\b_{g_5}$ and the
position of the $\SU(5)_\GUT$ Landau pole assuming $\alpha_5(\L_c) = 1/25$,
as would occur for $\L_c \sim M_\GUT$.  Also throughout this paper we
have assumed that there is an extra GUT-breaking adjoint in the
spectrum above $M_\GUT$, which gives a contribution to $\b_{g_5}$
in addition to that from the exotic sector and the usual matter fields.
This is a conservative assumption in that most realistic GUT-breaking
sectors will require at least this much additional matter.  However,
if a clever way could be found to combine the physics of GUT breaking and
conformal symmetry breaking, it is possible that the bounds could be somewhat
relaxed.

\begin{table}[t]
\begin{center}
\begin{tabular}{ c | c | c | c | c | c | c | c | c || c | c | c}
$N$ & $R_{T_1}$ & $R_{T_2}$ & $R_X$ & $R_{\bar Q}$ & $R_S$ & $R_{\bar X}$ & $R_Q$ & $R_{\bar S}$ & $\mathcal A$ & $\L_{\SU(5)}/\L_c$ & $\L/\L_c$\\
\hline
$6$ & $.740$ & $.706$ & $.625$ & $.616$ & $.673$ & $.621$ & $.635$ & $.616$ & $23.6$ & $10^{2.48}$ & $10^{22.91 \pm 4.33}$\\
$7$ & $.862$ & $.782$ & $.561$ & $.546$ & $.661$ & $.557$ & $.576$ & $.546$ & $32.6$ & $10^{1.80}$ & $10^{8.60 \pm 1.63}$\\
$8$ & $.992$ & $.885$ & $.497$ & $.483$ & $.620$ & $.495$ & $.511$ & $.483$ & $42.9$ & $10^{1.37}$ & $10^{4.96 \pm 0.77}$\\
$9$ & $1.123$ & $1.021$ & $.434$ & $.425$ & $.544$ & $.435$ & $.443$ & $.425$ & $54.4$ & $10^{1.08}$ & $10^{3.26 \pm 0.27}$\\
$10$ & $1.251$ & $1.196$ & $.373$ & $.369$ & $.429$ & $.375$ & $.377$ & $.369$ & $67.2$ & $10^{0.87}$ & $10^{2.35 \pm 0.01}$\\
\end{tabular}
\end{center}
\caption{$R$-charges in the $\tenrep + \fivebarrep + \trivrep$ model
with the superpotential $W = T_1 X Q + T_2 \bar{X} S$
assumed to be marginal.  The last three columns give: the
$\SU(5)_\GUT$ anomaly $\mathcal A$, the position of the
$\SU(5)_\GUT$ Landau pole assuming $\alpha_5(\L_c) = 1/25$, and the
phenomenologically required size for the conformal window.  We have assumed the presence of an additional GUT-breaking adjoint above $\L_c$.}
\label{tab:rchargetenfiveone}
\end{table}

In all cases, the required $\L$ is larger than the Landau pole
$\L_{\SU(5)}$.  One might try to avoid this fate by adding
additional relevant deformations to the superpotential.  For
instance, when $N \geq 8$, the operator $(\bar Q Q)^2$ is gauge
invariant and relevant, so one could impose the constraint that it
too becomes marginal.  However, we have not found a set of
relevant deformations that can save this model.

It's striking that such a simple model so badly violates
$\L<\L_{\SU(5)}$.  A natural question to ask is whether there
exist {\it any} simple vector-like models that can avoid this
bound. In Appendix~\ref{apx:listofmodels} we have listed all
qualitatively different simple group models containing two
distinct couplings to the $T_i$'s involving three or four fields.
Looking through this list, we find only two other models that
possess two three-field couplings.  The first is similar to the
present model and based on $\SU(N)$ with fundamentals in the
$\tenrep + \fiverep + \fivebarrep$ representation and
anti-fundamentals in the $\tenbarrep + \fivebarrep + \fiverep$
representation.  The second is based on $\Sp(2 N)$ with
fundamentals in the $\tenrep + \tenbarrep + \fiverep +
\fivebarrep$.  However, we find that both of these models have
similar problems with the $\SU(5)_{\GUT}$ Landau pole.  To go
forward, it is clear that we need to additionally consider models
with four-field couplings.

In fact, it appears likely that almost any vector-like CFT sector
containing chiral superfields $X$ in $\tenrep$'s and
$\tenbarrep$'s of $\SU(5)_\GUT$ will be problematic when one
considers $\b_{g_5}$.  The reason is that we typically need the
CFT to have a relatively large gauge group $G$ in order to have a
sufficiently strongly-coupled fixed point.  However, since the
fields $X$ transform in representations of $G$, they look like a
large number of SM $\tenrep + \tenbarrep$ pairs, each of which
contributes roughly a factor of $3$ (as opposed to $1$ for
$\fiverep + \fivebarrep$ pairs) to the anomaly coefficient
$\mathcal{A}$.  Further, strong gauge interactions tend to drive
$\g_X$ negative, which only increases the contribution to
$\b_{g_5}$.  These considerations motivate us to focus on CFT
sectors that contain only GUT $\fiverep$'s, $\fivebarrep$'s, and
singlets.  This cuts down the space of models considerably, and
there are only a few that we need to consider.

As summarized in Appendix~\ref{apx:listofmodels}, we find exactly
six possible model
structures.  Two of the models require baryonic operators that are
specific to $\SU(3)$ and $G_2$, respectively.  However, we
find that these models are too weakly coupled to be
phenomenologically successful and will not discuss them further.
The remaining models are based on $\SU(N)$ with an adjoint,
$\SO(N)$ with an adjoint or symmetric tensor, and $\Sp(2N)$ with
an anti-symmetric tensor.  We will consider these models and their
deformations below, and ultimately find that both the $\SU(N)$ models and the
$\Sp(2N)$ models can potentially be phenomenologically successful.

\subsection{\texorpdfstring{$\SU(N)$ with an Adjoint}{SU(N) with an Adjoint}}
\label{sec:sunadj}

Next we will consider perhaps the simplest candidate CFT based on
a strong $\SU(N)$ gauge group that does not contain any ${\bf
\bar{10}}$ representations of $\SU(5)_\GUT$.  In order to allow
for two distinct couplings to the Standard Model {\bf 10}'s, the
sector contains an $\SU(N)$ adjoint $A$ in addition to vector-like
pairs of fundamentals in the $\fiverep + \fivebarrep$
representation of $\SU(5)_\GUT$.  The matter content is given in
Table~\ref{tab:sunadjmodel}.  Without a superpotential, the
$\SU(N)$ theory is simply adjoint SQCD with $N_f=10$, which is
believed to flow to an interacting conformal fixed point for all
$0 < N_f < 2 N$~\cite{Kutasov:1995np,Kutasov:1995ss}, or equivalently $N \geq 6$.

\begin{table}[t]
\begin{center}
\begin{tabular}{ | l | c | c | }
\hline
                    & $\SU(5)_\GUT$ & $\SU(N)$ \\
\hline
$Q_1+\bar{Q}_2$   & $\fiverep + \fivebarrep$      & $\Yfund$ \\
$\bar{Q}_1+Q_2$   & $\fivebarrep + \fiverep$      & $\Yafund$ \\
$A$               & $\trivrep$              & Ad. \\
\hline
\end{tabular}
\end{center}
\caption{CFT sector based on $\SU(N)$ with an adjoint.}
\label{tab:sunadjmodel}
\end{table}

The two lowest dimension couplings to $\tenrep$'s are \beq
\label{eq:sunadjsuperpotential1} W_{\tenrep} &=& T_1 \bar Q_1 \bar Q_2
+ T_2 \bar Q_1 A \bar Q_2. \eeq As it stands, this theory doesn't
admit any other three- or four-field couplings to the Standard Model.  However, if desired it would be straightforward to
introduce additional couplings to $\fivebarrep$'s by adding extra
SM singlet flavors to the theory.

Assuming that the terms in \Eq{eq:sunadjsuperpotential1}
are the only marginal interactions, it is then straightforward to
determine the $\U(1)_R$ symmetry of this theory using
$a$-maximization.  The $R$-charges are constrained by the
superpotential and anomaly cancelation as
\bea\label{eq:conssunadjoint}
2 &=& R_{T_1} + R_{\bar{Q}_1} + R_{\bar{Q}_2} \nonumber\\
2 &=& R_{T_2} + R_{\bar{Q}_1} + R_A + R_{\bar{Q}_2} \\
0 &=& N + \frac52 (R_{Q_1} - 1) +  \frac52 (R_{Q_2} - 1) + \frac52
(R_{\bar{Q}_1} - 1) + \frac52 (R_{\bar{Q}_2} -1) + N (R_A -
1).\nonumber\eea Maximizing $a(R_t)$ subject to these constraints
then gives the $R$-charges in Table~\ref{tab:rchargesunadj}.  It
is important to note that $N \geq 15$ is required in order for the
second superpotential coupling to be marginal in the CFT.  This is
because for smaller values of $N$ we would find a violation of the
unitarity bound $R_{T_2} < 2/3$, indicating that the coupling must
flow to zero.  On the other hand, for all $N > 15$ the
gauge-invariant operator $\Tr[A^2]$ has $R < 2/3$, and we have
modified the $a$-maximization procedure as in
\Eq{eq:amodification} in order to account for this operator
becoming a free field.

\begin{table}[t]
\begin{center}
\begin{tabular}{ c | c | c | c | c | c || c | c | c }
$N$ & $R_{T_1}$ & $R_{T_2}$ & $R_{Q_{1,2}}$ & $R_{\bar Q_{1,2}}$ & $R_A$ & $\mathcal A$ & $\L_{\SU(5)}/\L_c$ & $\L/\L_c$\\
\hline
$15$ & $1.012$ & $.670$ & $.479$ & $.494$ & $.342$ & $41.6$ & $10^{1.41}$ & $-$\\
$16$ & $1.030$ & $.703$ & $.469$ & $.485$ & $.327$ & $45.4$ & $10^{1.29}$ & $-$\\
$17$ & $1.047$ & $.734$ & $.460$ & $.476$ & $.313$ & $49.2$ & $10^{1.19}$ & $-$\\
$18$ & $1.063$ & $.763$ & $.452$ & $.468$ & $.300$ & $53.1$ & $10^{1.10}$ & $-$\\
$19$ & $1.079$ & $.791$ & $.444$ & $.461$ & $.288$ & $57.0$ & $10^{1.03}$ & $-$\\
$20$ & $1.093$ & $.816$ & $.437$ & $.453$ & $.277$ & $61.0$ & $10^{0.96}$ & $10^{4.40 \pm 0.29}$\\
$21$ & $1.107$ & $.840$ & $.431$ & $.446$ & $.267$ & $65.0$ & $10^{0.90}$ & $10^{4.04 \pm 0.50}$\\
$22$ & $1.120$ & $.862$ & $.425$ & $.440$ & $.258$ & $69.0$ & $10^{0.85}$ & $10^{3.77 \pm 0.64}$\\
$23$ & $1.133$ & $.883$ & $.419$ & $.434$ & $.249$ & $73.1$ & $10^{0.80}$ & $10^{3.61 \pm 0.68}$\\
$24$ & $1.145$ & $.903$ & $.414$ & $.428$ & $.241$ & $77.2$ & $10^{0.76}$ & $10^{3.52 \pm 0.67}$\\
$25$ & $1.156$ & $.922$ & $.409$ & $.422$ & $.234$ & $81.3$ & $10^{0.72}$ & $10^{3.44 \pm 0.65}$\\
$26$ & $1.166$ & $.940$ & $.404$ & $.417$ & $.227$ & $85.5$ & $10^{0.69}$ & $10^{3.37 \pm 0.64}$\\
\end{tabular}
\end{center}
\caption{$R$-charges in the $\SU(N)$ adjoint model with
superpotential $W = T_1 \bar Q_1 \bar Q_2+T_2 \bar Q_1 A
\bar Q_2$ assumed to be marginal.  For $N > 15$ the operator
$\Tr[A^2]$ violates the unitarity bound and the effect of this
operator becoming a free field is included in the $a$-maximization
procedure. The last three columns give: the $\SU(5)_\GUT$ anomaly
$\mathcal A$, the position of the $\SU(5)_\GUT$ Landau pole
assuming $\alpha_5(\L_c) = 1/25$, and the phenomenologically required
size for the conformal window.}
\label{tab:rchargesunadj}
\end{table}

While we have arbitrarily stopped at $N = 26$, it is clear that in all
cases the phenomenologically required running distance can not be
achieved without hitting an $\SU(5)_{\GUT}$ Landau pole.  Thus, we
next consider whether deforming the theory by an additional
superpotential term can improve the situation.

\subsubsection{\texorpdfstring{Deformation by $\Tr[A^{k+1}]$}{Deformation by Tr[A^{(k+1)}]}}
Here we will consider what happens when we add $\Tr[A^{k+1}]$ as a marginal interaction in
the superpotential.  This deformation imposes the additional constraint that
$R_{A} = 2/(k+1)$.  Without the couplings to $T_{1,2}$, it is well known that this
theory has a dual description in terms of an $SU(k N_f - N)$ gauge group~\cite{Kutasov:1995ve,Kutasov:1995np,Kutasov:1995ss},
where here $N_f = 10$.  The dual gauge group becomes IR free when
\beq
N \geq (k-\frac12) N_f.
\eeq
We expect the same to be true in the theory with the couplings to $T_{1,2}$, since these simply
correspond to turning on mesonic mass operators in the dual description.  This gives an upper bound on the
values of $N$ for which this theory can flow to an interacting conformal fixed point.

First we will consider the $k=2$ deformation, namely adding $\Tr[A^3]$ as a marginal interaction.
In addition to needing $N < 15$ in order to flow to a non-trivial fixed point, we also find
that $R_{T_2} > 2/3$ requires $N > 10$.  Thus, we only need to consider a small range of $N$.
In Table~\ref{tab:rchargesunadja3} we give the result of $a$-maximization for this theory.\footnote{We are indebted to Nathaniel Craig for noticing that we mistakenly included incorrect values of $\mathcal A$ and $\L_{\SU(5)}/\L_c$ in a previous version of this table.  This led
us to prematurely conclude that the model was not viable.  For further discussion of this model and its variations,
see~\cite{Craig}.} For each $10 < N < 15$ we find that the operators
$Q_1 Q_{2}$, $Q_1 \bar{Q}_{1}$, $\bar{Q}_{2} Q_2$, and $\bar{Q}_{1} \bar{Q}_{2}$
have $R < 2/3$, and hence most of these operators must become free fields.  An important
subtlety is that the $\tenbarrep$ component of $\bar{Q}_{1} \bar{Q}_{2}$ is set to zero by the
$T_1$ equation of motion, and is not part of the chiral ring of the theory.
Because of this, the unitarity bound does not apply to this operator, and we should not
include it when modifying $a$ to account for the accidental symmetries.  In addition, as discussed in
Section~\ref{sec:amax}, when these operators become free one should modify not only the
$a$-maximization procedure but also the calculation of other anomalies.  We have been careful to
include these effects in the calculation of the $\SU(5)_\GUT$ anomaly $\mathcal A$.

\begin{table}[t]
\begin{center}
\begin{tabular}{ c | c | c | c | c | c || c | c | c }
$N$ & $R_{T_1}$ & $R_{T_2}$ & $R_{Q_{1,2}}$ & $R_{\bar Q_{1,2}}$ & $R_A$ & $\mathcal A$ & $\L_{\SU(5)}/\L_c$ & $\L/\L_c$\\
\hline
$11$ & $1.448$ & $.781$ & $.257$ & $.276$ & $.667$ & $33.884$ & $10^{1.73}$ & $-$\\
$12$ & $1.572$ & $.905$ & $.186$ & $.214$ & $.667$ & $34.528$ & $10^{1.70}$ & $-$\\
$13$ & $1.705$ & $1.039$ & $.119$ & $.147$ & $.667$ & $35.925$ & $10^{1.63}$ & $10^{1.79 \pm 0.14}$\\
$14$ & $1.849$ & $1.182$ & $.058$ & $.076$ & $.667$ & $38.080$ & $10^{1.54}$ & $10^{1.44 \pm 0.25}$\\
\end{tabular}
\end{center}
\caption{$R$-charges in the deformed $\SU(N)$ adjoint model with
superpotential $W = T_1 \bar Q_1 \bar Q_2+T_2 \bar Q_1 A
\bar Q_2 + \Tr[A^3]$ assumed to be marginal.  For these values of
$N$ the operators $Q_1 Q_{2}$, $Q_1 \bar{Q}_{1}$, $\bar{Q}_{2}
Q_2$, and the {\bf 15} component of $\bar{Q}_{1} \bar{Q}_{2}$ are
assumed to be free fields in the $a$-maximization procedure.  The
last three columns give: the $\SU(5)_\GUT$ anomaly $\mathcal A$,
the position of the $\SU(5)_\GUT$ Landau pole assuming $\alpha_5(\L_c)
= 1/25$, and the phenomenologically required size for the conformal
window.} \label{tab:rchargesunadja3}
\end{table}

Here we see that the situation is improved for $N=13,14$, since the
$\SU(5)_\GUT$ Landau pole can potentially occur at or above the top of the conformal
window, given our assumptions.  However, since $g_5$ is becoming fairly strongly
coupled at the top of the conformal window and the running distance is so short,
one might worry that our approximation of treating $\SU(5)_\GUT$ as a
flavor group is not very good.  The tension could be
eased somewhat if $\alpha_5(\L_c)$ were smaller than the unified value of
$1/25$, or if we could find a way to use composites of the CFT sector in order to break
the GUT group rather than introducing an additional $\SU(5)_{\GUT}$ adjoint
as we have assumed.  Nevertheless, we are motivated to see if we can find any
models where this tension can be avoided.

The next deformation to consider is $k=3$, or $\Tr[A^4]$.  Here we find that the
full superpotential can be interacting for $11 < N < 25$.  We give the result of $a$-maximization
for this theory in Table~\ref{tab:rchargesunadja4}, where again we have made sure that operators which
appear to violate the unitarity bound are treated as free fields.  We see that for $N > 20$ the $\SU(5)_\GUT$ Landau
pole can again occur just above the top of the required conformal window, so these models are potentially
viable.  For $N > 20$ we also see the appearance of negative $R$-charges, which may seem unusual at first
glance.  However, since we are taking care to make sure that there are no gauge-invariant operators which
violate the unitarity bound, we do not see any reason why these theories should be excluded.

\begin{table}[t]
\begin{center}
\begin{tabular}{ c | c | c | c | c | c || c | c | c }
$N$ & $R_{T_1}$ & $R_{T_2}$ & $R_{Q_{1,2}}$ & $R_{\bar Q_{1,2}}$ & $R_A$ & $\mathcal A$ & $\L_{\SU(5)}/\L_c$ & $\L/\L_c$\\
\hline
$12$ & $1.188$ & $.688$ & $.394$ & $.406$ & $.500$ & $37.762$ & $10^{1.55}$ & $-$\\
$13$ & $1.289$ & $.789$ & $.344$ & $.356$ & $.500$ & $44.350$ & $10^{1.32}$ & $-$\\
$14$ & $1.383$ & $.883$ & $.292$ & $.308$ & $.500$ & $47.826$ & $10^{1.22}$ & $-$\\
$15$ & $1.482$ & $.982$ & $.241$ & $.259$ & $.500$ & $50.080$ & $10^{1.17}$ & $10^{2.20 \pm 0.25}$\\
$16$ & $1.585$ & $1.085$ & $.192$ & $.208$ & $.500$ & $52.919$ & $10^{1.11}$ & $10^{1.83 \pm 0.35}$\\
$17$ & $1.690$ & $1.190$ & $.145$ & $.155$ & $.500$ & $56.345$ & $10^{1.04}$ & $10^{1.64 \pm 0.31}$\\
$18$ & $1.797$ & $1.297$ & $.099$ & $.101$ & $.500$ & $60.362$ & $10^{0.97}$ & $10^{1.49 \pm 0.28}$\\
$19$ & $1.900$ & $1.400$ & $.050$ & $.050$ & $.500$ & $61.300$ & $10^{0.96}$ & $10^{1.36 \pm 0.26}$\\
$20$ & $2.000$ & $1.500$ & $.000$ & $.000$ & $.500$ & $61.000$ & $10^{0.96}$ & $10^{1.26 \pm 0.24}$\\
$21$ & $2.100$ & $1.600$ & $-.050$ & $-.050$ & $.500$ & $61.300$ & $10^{0.96}$ & $10^{1.15 \pm 0.19}$\\
$22$ & $2.200$ & $1.700$ & $-.100$ & $-.100$ & $.500$ & $62.200$ & $10^{0.94}$ & $10^{1.05 \pm 0.16}$\\
$23$ & $2.300$ & $1.800$ & $-.150$ & $-.150$ & $.500$ & $63.700$ & $10^{0.92}$ & $10^{0.97 \pm 0.13}$\\
$24$ & $2.400$ & $1.900$ & $-.200$ & $-.200$ & $.500$ & $61.800$ & $10^{0.95}$ & $10^{0.90 \pm 0.11}$\\
\end{tabular}
\end{center}
\caption{$R$-charges in the deformed $\SU(N)$ adjoint model with
superpotential $W = T_1 \bar Q_1 \bar Q_2+T_2 \bar Q_1 A
\bar Q_2 + \Tr[A^4]$ assumed to be marginal.  For $N > 13$ the operators
$Q_1 Q_{2}$, $Q_1 \bar{Q}_{1}$, $\bar{Q}_{2}Q_2$, and the
{\bf 15} component of $\bar{Q}_{1} \bar{Q}_{2}$ are
assumed to be free fields in the $a$-maximization procedure.  For $N > 17$ the
operators $Q_1 A Q_{2}$, $Q_1 A \bar{Q}_{1}$, $\bar{Q}_{2} A Q_2$, and the
{\bf 15} component of $\bar{Q}_{1} A \bar{Q}_{2}$ are
additionally assumed to be free fields.  Finally, for $N = 24$ the operators
$Q_1 A^2 Q_{2}$, $Q_1 A^2 \bar{Q}_{1}$, $\bar{Q}_{2} A^2 Q_2$, and
$\bar{Q}_{1} A^2 \bar{Q}_{2}$ are also free.  The
last three columns give: the $\SU(5)_\GUT$ anomaly $\mathcal A$,
the position of the $\SU(5)_\GUT$ Landau pole assuming $\alpha_5(\L_c)
= 1/25$, and the phenomenologically required size for the conformal
window.} \label{tab:rchargesunadja4}
\end{table}

In all of these cases, however, $g_5$ becomes fairly strongly coupled towards the top
of the conformal window, and it is not clear to what extent our approximations can be trusted.  The root
of the problem is that there is simply still too much matter in the CFT sector charged under $\SU(5)_{\GUT}$
relative to how strongly coupled the theory is.  Thus, we are motivated to try to find CFT sectors that are
even more efficient -- we wish to minimize the matter content of the CFT in order to stay deep
within the conformal window while still allowing two couplings to Standard Model $\tenrep$'s.

There are two remaining classes of vector-like models that do not
contain GUT $\tenrep$'s that we need to consider.  The first is
based on $\SO(N)$ with an adjoint (or a symmetric tensor) and two
fundamental $\fiverep + \fivebarrep$ pairs, which can be thought
of as simply taking a subgroup of the present model.  However, it
is easy to see that this model will be significantly worse than
the $\SU(N)$ version.  The reason is that an adjoint of $\SO(N)$
has Dynkin index $N-2$, which scales like $N$ just as in $\SU(N)$.
However, fundamentals of $\SO(N)$ have index $1$ rather than $1/2$
for $\SU(N)$, and since there are the same number of fundamentals
as in the $\SU(N)$ model $N$ will need to be roughly twice as
large to get to the same part of the conformal window.  The
contribution to the $\SU(5)_{\GUT}$ $\beta$-function is then
roughly twice as large, and the model is quickly ruled out.

The final model is based on $\Sp(2 N)$ with an anti-symmetric
tensor and fundamentals in the $\fiverep + \fivebarrep$
representation.  The biggest advantage of this model is that it
only needs half as many fundamentals in order to introduce
couplings to the $T_i$'s, and is hence a good candidate in our
search for a more efficient model.  In the following section we
will proceed to study this model and its possible deformations in
more detail.

\subsection{\texorpdfstring{$\Sp(2 N)$ with an Anti-symmetric Tensor}{Sp(2N) with an Anti-symmetric Tensor}}
\label{sec:spnanti}

The matter content of the $\Sp(2N)$ model is summarized in Table~\ref{tab:spnmodel}.
In order for the theory to be IR-interacting, we must have $N\geq 4$~\cite{Intriligator:1995ff}.
The two lowest dimension couplings to the Standard Model
are\beq \label{eq:sp2nadjcouplings} W_\tenrep
&=& T_1 \bar Q \bar Q+T_2 \bar Q A\bar Q. \eeq  There are no
three- or four-field couplings to $\fivebarrep$'s, though we will
later consider an extension of this model which can couple to
$\bar H$ and $\bar F_i$.

\begin{table}[h]
\begin{center}
\begin{tabular}{ | l | c | c | }
\hline
                    & $\SU(5)_\GUT$ & $\Sp(2 N)$ \\
\hline
$Q+\bar{Q}$       & $\fiverep + \fivebarrep$      & $\Yfund$ \\
$A$               & $\trivrep$              & $\Yasymm$ \\
\hline
\end{tabular}
\end{center}
\caption{Matter content of the $\Sp(2N)$ model.}
\label{tab:spnmodel}
\end{table}

Now we will determine the superconformal $\U(1)_R$
symmetry, assuming $W_{\tenrep}$ contains the only marginal
interactions.  The superpotential and anomaly cancelation impose
the constraints \beq
2 &=& R_{T_1}+2R_{\bar Q}\nonumber\\
2 &=& R_{T_2}+2R_{\bar Q}+R_A\\
0 &=& 2(N+1)+5(R_Q-1)+5(R_{\bar Q}-1)+2(N-1)(R_A-1).\nonumber \eeq
Performing $a$-maximization gives the $R$-charges listed in
Table~\ref{tab:rchargespn}, where we have arbitrarily stopped at $N=10$.
This model can evade the bound $\L<\L_{\SU(5)}$ when $N=5,6,7$ and $8$ (but the constraint
is too strong at larger values).  However, the required running
distance in all of these cases is $\L / \L_c \gsim M_{\pl} /
M_\GUT \sim 10^{2\mathrm{\ or\ }3} $.  It is not necessarily fatal
to have the upper end of the conformal window near $M_\pl$.
However if $\L\gsim M_\pl$, we lose confidence in our na\"\i ve
calculation of Standard Model wave-function factors, since
Planck-scale matter will likely influence the anomalous
dimensions.  The case of $N=8$ is perhaps the best behaved in this
light.

\begin{table}[t]
\begin{center}
\begin{tabular}{ c | c | c | c | c | c || c | c | c}
$N$ & $R_{T_1}$ & $R_{T_2}$ & $R_{Q}$ & $R_{\bar Q}$ & $R_A$ & $\mathcal A$ & $\L_{\SU(5)}/\L_c$ & $\L/\L_c$\\
\hline
$4$ & $1.045$ & $.778$ & $.401$ & $.477$ & $.268$ & $8.255$ & $10^{7.09}$ & $-$\\
$5$ & $1.103$ & $.872$ & $.382$ & $.448$ & $.231$ & $11.662$ & $10^{5.02}$ & $10^{3.85 \pm 0.73}$\\
$6$ & $1.154$ & $.950$ & $.369$ & $.423$ & $.204$ & $15.277$ & $10^{3.83}$ & $10^{3.45 \pm 0.65}$\\
$7$ & $1.197$ & $1.014$ & $.359$ & $.401$ & $.183$ & $19.076$ & $10^{3.07}$ & $10^{3.09 \pm 0.51}$\\
$8$ & $1.234$ & $1.067$ & $.351$ & $.383$ & $.166$ & $23.025$ & $10^{2.54}$ & $10^{2.76 \pm 0.34}$\\
$9$ & $1.263$ & $1.111$ & $.344$ & $.368$ & $.152$ & $27.076$ & $10^{2.16}$ & $10^{2.55 \pm 0.26}$\\
$10$ & $1.288$ & $1.147$ & $.338$ & $.356$ & $.140$ & $31.215$ & $10^{1.88}$ & $10^{2.40 \pm 0.20}$ \\
\end{tabular}
\end{center}
\caption{$R$-charges in the $\Sp(2N)$ model with the superpotential
$W_\tenrep = T_1 \bar{Q}\bar{Q}+T_2\bar{Q} A \bar{Q}$ assumed to be
marginal.  When the operators $\Tr[A^k]$ violate the unitarity
bound they are assumed to become free fields and the resulting
accidental symmetry is included in the $a$-maximization procedure.}
\label{tab:rchargespn}
\end{table}

\subsubsection{\texorpdfstring{Deformation by $\Tr[A^{k+1}]$}{Deformation by Tr[A^{(k+1)}]}}

Next, we will analyze possible deformations of our CFT, some of
which improve consistency with the bounds $\L<\L_{\SU(5)}$ and
$\L<M_\pl$.  First, we can consider adding $\Tr[A^{k+1}]$ to the
superpotential as a marginal interaction.  Note that the theory without
$W_\tenrep$ has a known dual description with gauge group
$\Sp(2k(N_f-2)-2N)$~\cite{Intriligator:1995ff}, where $N_f=5$
in our case.  This dual magnetic theory is IR free when
\beq\label{eq:spnAkbound} N > (k-\frac12)N_f - 2(k-1)=3k-\frac 1 2.\eeq
We expect the same to be true in the theory with $W_\tenrep$ turned on,
since in the dual description (which we will discuss further below) it
simply corresponds to deforming the theory by mesonic mass operators.
Consequently, our $\Tr[A^{k+1}]$ deformation engenders an upper limit
on values of $N$ for which the theory can have a non-trivial conformal fixed point.

Let us start by considering the theory deformed by $\Tr[A^3]$, which
can have a non-trivial conformal fixed point when $N=4$ or $5$.
Performing $a$-maximization gives the $R$-charges in
Table~\ref{tab:rchargespnA3}.  In the $a$-maximization procedure, we
have been careful to take into account the accidental symmetries
associated with the operators $QQ$ and $\bar Q Q$ becoming free
fields. (Note that we need not include $\bar{Q} \bar{Q}$, as it is
zero in the chiral ring and the unitarity bound does not apply.) As
discussed in Section~\ref{sec:amax}, this modifies not only the
$a$-maximization procedure, but also the calculation of other
anomalies, in particular the numerator of $\b_{g_5}$.  As we can
see from Table~\ref{tab:rchargespnA3}, this is a dramatic effect.  The
bounds from the Landau pole $\L_{\SU(5)}$ become much weaker, because
the $\{L,M\}$ pairs associated to $QQ$ and $\bar Q Q$ give large
negative contributions to $\mathcal A$.  The theory with $N=5$ is seen to easily
evade the Landau pole constraint and fits beautifully between
$M_{\GUT}$ and $M_{\pl}$.

\begin{table}[t]
\begin{center}
\begin{tabular}{ c | c | c | c | c | c || c | c | c}
$N$ & $R_{T_1}$ & $R_{T_2}$ & $R_{Q}$ & $R_{\bar Q}$ & $R_A$ & $\mathcal A$ & $\L_{\SU(5)}/\L_c$ & $\L/\L_c$\\
\hline
$4$ & $1.497$ & $.830$ & $.149$ & $.251$ & $.667$ & $6.063$ & $10^{9.66}$ & $-$\\
$5$ & $1.786$ & $1.119$ & $.026$ & $.107$ & $.667$ & $7.163$ & $10^{8.18}$ & $10^{1.57 \pm 0.22}$\\
\end{tabular}
\end{center}
\caption{$R$-charges in the $\Sp(2N)$ model with the superpotential
$W = T_1 \bar{Q}\bar{Q}+T_2\bar{Q} A \bar{Q}+\Tr[A^3]$.
assumed to be marginal.  Note that the $QQ$ and $\bar Q Q$ operators becoming free results in a significant negative contribution to $\mathcal A$.}
\label{tab:rchargespnA3}
\end{table}

Next let us consider the electric theory with $\Tr[A^4]$ as a marginal
interaction, which allows a non-trivial fixed point for $4 \leq N \leq
8$.  The results of $a$-maximization are summarized in
Table~\ref{tab:rchargespnA4}.  We find that the models with $N =
5,6,7,$ and $8$ can also be phenomenologically successful.  When the
operators $Q Q$, $\bar Q Q$, $Q A Q$, and $\bar Q A Q$ violate the
unitarity bound they are assumed to become free fields.  Note that we
need not do the same for $\bar Q \bar Q$ and $\bar Q A \bar Q$ because
they are set to zero in the chiral ring and the unitarity bound does
not apply.  For $N=8$ we once again see the appearance of negative $R$-charges.
However, we do not see any obvious reason why this theory should be excluded.

We could continue and classify viable models with $k\geq 5$.  However,
an important point is that the existence of both couplings $T_1 \bar Q
\bar Q$ and $T_2 \bar Q A\bar Q$ in tandem disallows any non-trivial
flavor symmetry for $A$.  Thus, all couplings $\Tr[A^{k+1}]$ are
necessarily allowed, and it's perhaps unnatural to expect that our
theory should flow to a fixed point with, for instance, $\Tr[A^5]$ as
a marginal operator instead of $\Tr[A^3]$.  At this point we could
also consider deformations by operators like $\bar Q Q \bar Q Q$, or
$\bar Q A^2 Q$.  However, we do not find that these lead to a
successful phenomenology, and so will not discuss these deformations
in detail.

\begin{table}[t]
\begin{center}
\begin{tabular}{ c | c | c | c | c | c || c | c | c}
$N$ & $R_{T_1}$ & $R_{T_2}$ & $R_{Q}$ & $R_{\bar Q}$ & $R_A$ & $\mathcal A$ & $\L_{\SU(5)}/\L_c$ & $\L/\L_c$\\
\hline
$4$ & $1.331$ & $.831$ & $.266$ & $.334$ & $.500$ & $8.460$ & $10^{6.92}$ & $-$ \\
$5$ & $1.531$ & $1.031$ & $.166$ & $.234$ & $.500$ & $9.960$ & $10^{5.88}$ & $10^{2.00 \pm 0.32}$ \\
$6$ & $1.787$ & $1.287$ & $.093$ & $.107$ & $.500$ & $12.409$ & $10^{4.72}$ & $10^{1.50 \pm 0.28}$ \\
$7$ & $2.000$ & $1.500$ & $.000$ & $.000$ & $.500$ & $13.000$ & $10^{4.64}$ & $10^{1.26 \pm 0.23}$\\
$8$ & $2.200$ & $1.700$ & $-.100$ & $-.100$ & $.500$ & $14.200$ & $10^{4.24}$ & $10^{1.05 \pm 0.16}$
\vspace{.2in}
\end{tabular}
\end{center}
\caption{$R$-charges in the $\Sp(2N)$ model with the superpotential
$W = T_1 \bar{Q}\bar{Q}+T_2\bar{Q} A \bar{Q}+\Tr[A^4]$ assumed to be marginal.  Note that when the $QQ$, $\bar Q Q$, $Q A Q$, and $\bar Q A Q$ operators become free they result in a significant negative contribution to $\mathcal A$.}
\label{tab:rchargespnA4}
\end{table}

\subsubsection{Dual Description}

Here we will briefly discuss the dual
description~\cite{Intriligator:1995ff} of the theories with a
$\Tr[A^{k+1}]$ superpotential.  The matter content of the dual is
summarized in Table~\ref{tab:spnmodeldual}, and the interacting superpotential
is given by
\beq
\label{eq:sp2ndualW}
W_{dual} &=& \Tr\, Y^{k+1}+\sum_{j=1}^k \p{M^j_{QQ}\bar q Y^{k-j} \bar q+M^j_{\bar QQ}q Y^{k-j} \bar q+M^j_{\bar Q\bar Q}q Y^{k-j} q}.
\eeq

\begin{table}[h]
\begin{center}
\begin{tabular}{ | l | c | c | }
\hline
                    & $\SU(5)_\GUT$ & $\Sp(6k-2N)$ \\
\hline
$q+\bar{q}$       & $\fiverep + \fivebarrep$      & $\Yfund$ \\
$Y$               & $\trivrep$              & $\Yasymm$ \\
$M^j_{QQ}+M^j_{\bar Q Q}+M^j_{\bar Q\bar Q}$ & $\tenrep+(\mathbf{24}+\trivrep)+\tenbarrep$ & $\trivrep$\\
\hline
\end{tabular}
\end{center}
\caption{Matter content of the dual description of the $\Sp(2N)$ model
  deformed by $\Tr[A^{k+1}]$.  The mesons $M^j_{QQ}$ ($j=1,\dots,k$)
  correspond to the operators $Q A^{j-1}Q$ in the electric
  theory. Similarly, $M^j_{\bar Q Q}\sim \bar Q A^{j-1}Q$ and
  $M^j_{\bar Q\bar Q}\sim \bar Q A^{j-1}\bar Q$.}
\label{tab:spnmodeldual}
\end{table}

After adding $W_\tenrep=T_1 M^1_{\bar Q\bar Q} + T_2 M^2_{\bar Q\bar Q}$
to the magnetic theory, we could consider integrating out the
massive pairs $\{T_1,M^1_{\bar Q\bar Q}\}$ and $\{T_2, M^2_{\bar Q\bar
  Q}\}$ to get a description with fewer degrees of freedom.  However,
including the relevant deformations $W_{exit} = \Tr M^1_{\bar Q Q} +
\Tr Y^2$ and expanding the theory around its supersymmetric vacua will
induce additional mass operators for the mesons, and in general there
will be linear combinations of $T_{1,2}$ with $M^j_{Q Q}$ that get
lifted.  This can introduce additional (incalculable) mixing angles
into the Yukawa couplings, which we are here assuming are $O(1)$.  We
present further discussion of these issues in Appendix~\ref{sec:CFTexit}.

It is also simple to describe operators becoming free in the dual
description.  If the full superpotential \Eq{eq:sp2ndualW} were
marginal, we would find that the gauge singlet operators
$M^1_{QQ}$ and $M^1_{\bar QQ}$ violate the unitarity bound.  Since
their $R$-charges cannot drop below $2/3$, the couplings
$M^1_{QQ}\bar q Y^{k-1} \bar q$ and $M^1_{\bar QQ}q Y^{k-1} \bar q$
must become irrelevant and flow to zero.  Similar considerations
apply to the couplings of $M^2_{QQ}$ and $M^2_{\bar QQ}$ for $N = 7,8$
in the case of $k=3$.  Leaving out these couplings and
performing $a$-maximization in the dual theory gives $R$-charges and
$\mathcal A$ in agreement with Tables~\ref{tab:rchargespnA3} and~\ref{tab:rchargespnA4}.
From this point of view the small values of $\mathcal A$ are not
surprising, since the theory has only a few GUT multiplets, and many
of their $R$-charges are at or near their free values.

\subsubsection{\texorpdfstring{Coupling to $\fivebarrep$'s}{Coupling to 5bar's}}

As reviewed in Section~\ref{sec:flavor}, a $\tenrep$-centered model
(with suppression factors only for $T_1$ and $T_2$) is possible at
large $\tan \b$, perhaps given some lucky $O(1)$ factors.  If we want
a superconformal flavor model that works at small $\tan\b$, we should
additionally generate suppression factors for $\fivebarrep$'s.  The
tradeoff between $\tan\b$ and $\e_{\bar F_i, \bar H}$ is expressed in
Eqs.~(\ref{eps5quarks}) and~(\ref{eps5leptons}).

Without modifying our theory, the lowest dimension coupling to a
Standard Model $\fivebarrep$ is $\bar Q^4 \bar H$ or $\bar Q^4
\bar F_i$.  However, $a$-maximization quickly rules this out as a
possible marginal interaction, since $\bar H$ or $\bar F_i$ would
be forced to have $R$-charge less than $2/3$, in violation of
unitarity.  An alternative approach is to add a pair of GUT
singlets $S+S'$ transforming as $(1,\Yfund)+(1,\Yfund)$ under
$\SU(5)_\GUT\x \Sp(2N)$.\footnote{Note that the $\Sp(2N)$ global anomaly forces us to add
fundamentals in pairs.} The theory now admits couplings \beq
W_{\fivebarrep}&=&\bar F_1QS + \bar H Q S'. \eeq To avoid
problematic dimension 3 and 4 baryon- and lepton-number violating
operators (like $H\bar F_i$ and $T_i \bar{F}_j \bar{F}_k$), we can impose any
$\U(1)$ flavor symmetry in the family shown in
Table~\ref{tab:flavorsymm}.  This family is uniquely determined by
the requirement of consistency with $W_{\tenrep}$, the SM Yukawas,
and either of the interactions $\bar F_i Q S$ or $\bar H Q S'$. In
particular, it's impossible to use a $\U(1)$ flavor symmetry to
allow one of the interactions in $W_{\fivebarrep}$ and disallow
the other.

\begin{table}[t]
\begin{center}
\begin{tabular}{ c | c | c | c | c | c | c | c | c | c}
& $T_i$ & $\bar F_i$ & $H$ & $\bar H$ & $Q$ & $\bar Q$ & $S$ & $S'$ & $A$\\
\hline
$F$ & 1 & $x$  & $-2$ & $-1-x$ & 1/2 & $-1/2$ & $-x-1/2$ & $x+1/2$ & 0
\end{tabular}
\end{center}
\caption{Possible $\U(1)$ flavor charges that disallow
low-dimension baryon- and lepton-number violating operators.
The parameter $x$ is arbitrary.} \label{tab:flavorsymm}
\end{table}

Without doing anything quantitative, we can anticipate that a model
with marginal interactions $W_\tenrep+W_{\fivebarrep}$ will give
anomalous dimensions for $\bar H$ and $\bar F_i$ that are comparable
to the anomalous dimension of $T_1$, since both come from three-field
interactions with strong-sector mesons.  This will produce $\e_{\bar
  H}$ and $\e_{\bar F_1}$ factors that are unacceptably small if
$W_{\fivebarrep}$ remains marginal throughout the range
$\mu\in[\L_c,\L]$.  One possible resolution is that the pair $S+S'$
develops a large mass and decouples above $\L_c$, so that the
anomalous dimensions $\g_{\bar F_i}$ and $\g_{\bar H}$ are
exponentiated over a shorter running distance than $\g_{T_1}$.  This
then introduces the decoupling scale of $S,S'$ as a new parameter into
our theory, which diminishes the theory's predictivity.

However, one prediction that remains follows from the ${\mathbb Z}_2$
symmetry relating $(\bar F_1,S)$ and $(\bar H, S')$.  We see that
$R_{\bar F_1}=R_{\bar H}$, so $\e_{\bar F_1}=\e_{\bar H}$, regardless
of the decoupling scale for $S,S'$.  In principle, if we could compute
the $O(1)$ mixing angles that enter Yukawa couplings when exiting the
conformal window (in particular, if we knew the origin of the
violation of GUT mass relations), then this symmetry combined with
Eqs.~(\ref{eps5quarks}) and~(\ref{eps5leptons}) would yield a
prediction for $\tan \b$.

\section{A Previously `Incalculable' Model}
\label{sec:incalculable}

Several more examples of superconformal flavor models were presented
in the initial paper on the subject~\cite{Nelson:2000sn}.  For some of
them, the $R$-charges could be determined uniquely from superpotential
constraints and anomaly cancelation alone.  However, such models
typically involve a complicated superpotential with lots of marginal
operators in order to fully constrain the space of trial $R$-charges.
By contrast, models with simpler superpotentials were `incalculable'
at the time~\cite{Nelson:2000sn} was written, and the task of
determining their viability as flavor models was left for future work.
In this section, we will apply $a$-maximization to determine the
$R$-charges for one such `incalculable' model.

The model we consider has two exotic $\Sp(8)$ gauge groups, which we
will refer to as $\Sp(8)$ and $\Sp(8)'$.  The matter content is given
in Table~\ref{tab:incalculableten}.  There are $N_f = 9$ pairs of
fundamentals under $\Sp(8)$ ($N_c = 4$).  In the absence of a
superpotential, and ignoring the $\Sp(8)'$ coupling, this gauge group
enters a non-Abelian Coulomb phase because $3/2 (N_c + 1) < N_f < 3
(N_c+ 1)$~\cite{Intriligator:1995ne}.  In addition, there are $N_f'=6$
pairs of fundamentals in a confining $\Sp(8)'$, which are introduced
in order to allow all exotic states to decouple from the low-energy
spectrum.

\begin{table}[h]
\begin{center}
\begin{tabular}{ | l | c | c | c | }
\hline
                    & $\SU(5)_\GUT$ & $\Sp(8)$ & $\Sp(8)'$ \\
\hline
$Q$                    & $\tenbarrep$          & $\Yfund$ & $1$\\
$L,M,J_{1...6}$         & $1$                 & $\Yfund$  & $1$\\
$\bar{Q}'$             & $\tenrep$                & $1$  & $\Yfund$\\
$\bar{J}_{1,2}^{'}$     & $1$                 & $1$  & $\Yfund$\\
\hline
\end{tabular}
\end{center}
\caption{Matter content of the `{\bf 10}-centered' model presented
in~\cite{Nelson:2000sn}.} \label{tab:incalculableten}
\end{table}

The theory is assumed to flow to a fixed point with the marginal
interactions \beq W_{int} = (J_1 J_2)^2 + (J_3 J_4)^2 + (J_5
J_6)^2 + (L J_1) (J_1 J_3) + T_2 Q M + T_1 Q L. \eeq While there
is much that could be said about the structure of flows that could
lead to this fixed point, as well as exit from the CFT regime and
the decoupling of exotic states, here we will simply demonstrate
that this model cannot be realistic due to the $R$-charges alone.

The superpotential and anomaly cancelation impose the constraints
\bea\label{eq:consincalculableten}
2 &=& R_{T_1} + R_Q + R_L \\
2 &=& R_{T_2} + R_Q + R_M \nonumber\\
2 &=& 2 R_{J_1} + 2 R_{J_2} \nonumber\\
2 &=& 2 R_{J_3} + 2 R_{J_4} \nonumber\\
2 &=& 2 R_{J_5} + 2 R_{J_6} \nonumber\\
2 &=& 2 R_{J_1} + R_{J_3} + R_{L} \nonumber\\
0 &=& 2 (N_c+1) + 10 (R_Q-1) + (R_L-1) + (R_M-1) + \sum_{i=1}^6 (R_{J_i}-1).\nonumber\eea
Maximizing $a(R_t)$ subject to these constraints then yields the
$R$-charges given in Table~\ref{tab:rchargeincalculableten}.  From
these results we can see immediately that generating a hierarchy is
not possible due to the $R$-charges for $T_1$ and $T_2$ being
approximately equal.  Thus, we conclude that this model is not
viable.\footnote{By contrast, the calculable examples
  in~\cite{Nelson:2000sn} do successfully generate a hierarchy over a
  small range of scales. However, we should note that they are still
  subject to the Landau pole constraint $\L_{\SU(5)}<\L$.  The
  `\tenrep-centered model without proton decay'
  in~\cite{Nelson:2000sn} develops an $\SU(5)_\GUT$ (or $\SU(3)$) Landau pole after $\sim 3.5$ decades of running, so cannot work down to 10 TeV, as the authors claim.}

\begin{table}[t]
\begin{center}
\begin{tabular}{ c | c | c | c | c | c | c | c | c | c | c }
$R_{T_1}$  & $R_{T_2}$   & $R_Q$ & $R_L$ & $R_M$ & $R_{J_1}$ & $R_{J_2}$ & $R_{J_3}$ & $R_{J_4}$ & $R_{J_5}$ & $R_{J_6}$  \\
\hline
1.104     &    1.107   &  .401 & .494  & .491  &  .502  &  .498  & .501     & .499      &  .500    &  .500  \\
\end{tabular}
\end{center}
\caption{$R$-charges in the `$\tenrep$-centered' model presented in~\cite{Nelson:2000sn}.}
\label{tab:rchargeincalculableten}
\end{table}

\section{Conclusions}
\label{sec:concl}

It is not hard to imagine that the spectrum just below $M_\pl$
includes an exotic non-Abelian gauge group $G$ and vector-like matter
charged under both $G$ and $\SU(5)_\GUT$.  Often such sectors are
assumed to be lifted at a high scale and ignored for the sake of
low-energy physics.  But we've seen that some very simple exotic
sectors can naturally generate flavor hierarchies in the Standard Model,
dynamically distinguishing the different generations in a way
determined simply by representation theory.  For model builders
working at the Planck scale, these kinds of exotic sectors are
important to keep in mind as viable and well-motivated extensions to
the MSSM.

Given this, it is worthwhile to investigate precisely which exotic
sectors are phenomenologically interesting and viable.  Luckily,
$a$-maximization allows us to quantitatively evaluate a large class of
superconformal flavor models that would be otherwise incalculable.
Vector-like theories are a good starting point for model building
because it's easy to ensure by adding mass terms that conformal
symmetry is broken and all exotic states are lifted from the
low-energy spectrum.  However, we have seen that they are also highly
constrained by demanding that the $\SU(5)_\GUT$ coupling remain
perturbative over the range of energies required to produce a
reasonable Yukawa hierarchy.  Roughly, a large hierarchy requires
large anomalous dimensions, and therefore a strongly-coupled SCFT.
But strong coupling usually requires a large exotic gauge group, and
therefore forces us to include many GUT multiplets, driving the GUT
$\b$-function highly positive.

In this paper, we focused on possibly the simplest set of vector-like
models: $\tenrep$-centered models with a simple gauge group and a
small number of superpotential operators.  Among these, only models
based on $\Sp(2N)$ with an anti-symmetric tensor seem to be efficient enough
to easily evade the GUT Landau pole constraint.  Some models based on
$\SU(N)$ with an adjoint are also potentially viable.  However, there
are a few obvious generalizations that might yield other viable superconformal
flavor models.  Firstly, it would be interesting to do a systematic
study of chiral exotic sectors.  These might be better able to avoid
the Landau pole constraint because large GUT multiplets needn't come
in vector-like pairs.  However, ensuring that all exotic states
decouple is clearly a more delicate issue.  Secondly, one might try
finding models where a large number of composite operators in
non-trivial GUT multiplets become free, in order to exploit the
negative contribution to $\b_{g_5}$ from $\{L,M\}$ pairs discussed
in Sections~\ref{sec:amax} and~\ref{sec:spnanti}.  Alternatively, one
could investigate extending the exotic sector to include fields that
don't couple directly to $T_1$ or $T_2$ in the superpotential, but
affect their anomalous dimensions through other marginal
superpotential couplings (as in~\cite{Nelson:2000sn}).  Further, one
might generalize to non-simple gauge groups, or include additional
$U(1)$ or discrete symmetries to constrain the allowed marginal
operators.

We suspect that in each of these generalized classes of models, the
Landau pole constraint allows only a few possibilities.  Meanwhile, in
non-GUT models, requiring that the $\SU(3)$ coupling remain
perturbative should provide an even stronger constraint, since
$\a_{strong}$ is strictly greater than $\a_\GUT$ and we lose the
negative $\b$-function contribution from $XY$ gauge bosons.  It seems
unlikely that a flavor CFT could exist down to the TeV scale unless we
are willing to give up on the idea of perturbative unification.

Superconformal flavor models are related via AdS/CFT to 5D warped
flavor models, which generate Yukawa hierarchies through sequestering
in an extra dimension.  In the dual picture to our setup, the Higgs,
$T_3$, and other fields decoupled from the exotic sector are localized
in the UV,\footnote{Note that we are relying on SUSY to solve the
  hierarchy problem, so there is no need to put the Higgs in the IR.}
while $T_1$ and $T_2$ correspond to massive fields localized near the
IR.  Warped flavor models are usually analyzed in the 5D supergravity
limit, which corresponds to a large-$N$ limit in the dual CFT.
However, large-$N$ CFTs are precisely those for which the Landau pole
constraint is strongest, and one might worry that this places a severe
restriction on calculable warped flavor models.  Certainly, there seem
to be only a limited number of viable superconformal flavor models
even at small-$N$, and it would be interesting to better understand
the nature of these limitations in a supergravity description.

It is also suggestive that viable superconformal flavor models can fit
nicely below $M_\pl$, with conformal symmetry broken near $M_\GUT$.
The exotic sectors we write down are similar to previously considered
models of GUT
breaking~\cite{Kitano:2005ez,Nomura:2006pn,Kitano:2006wm}, and it
would be interesting to try to combine the physics of conformal
symmetry breaking and GUT breaking (and perhaps even SUSY breaking) in
some way.  To this end, it would be good to better understand the
vacuum structure of the more realistic models in the presence of
various relevant deformations.  Further, after combining these flavor
models with a more realistic picture of GUT physics and SUSY breaking,
it would be interesting to see if deviations from GUT mass relations
in the lighter generations can be accommodated at a more quantitative
level.

Finally, we would like to stress that without a full understanding of
the origin of the Yukawa hierarchies, one doesn't know the extent to
which the `SUSY flavor problem' is really a problem.  In the present
scenario, scalar mass operators which are potentially flavor-violating
at a high scale will also receive suppressions by the CFT
dynamics~\cite{Nelson:2001mq,Kobayashi:2001kz}.  It would be
interesting, for example, to extend the analysis
of~\cite{Nomura:2007ap,Nomura:2008gg} on flavor constraints to the
present case, where SUSY-breaking operators may receive extra suppressions due
to having different anomalous dimensions under the strong dynamics.
This may then help guide us to a more coherent picture of what
low-scale measurements can tell us about the relationship between
flavor dynamics and the mediation of supersymmetry breaking.  If we
are lucky, the mechanism underlying the flavor hierarchies will then
leave its imprint at the LHC, and measurements of the superpartner
spectrum will help us to come several steps closer to unraveling the
deep mysteries of nature.

\section*{Acknowledgements}

We thank Clay Cordova, Nathaniel Craig, Yanou Cui, Ken Intriligator, Andrey Katz, John Mason,
David Morrissey, Lisa Randall, Matt Schwartz, and Brian Wecht for helpful
comments and conversations.  This work is supported in part by the Harvard
Center for the Fundamental Laws of Nature and by NSF grant PHY-0556111.
DP would like to thank the Galileo Galilei Institute for its hospitality
during the completion of this work.

\appendix

\section{Sketch of a Proof of $a$-Maximization}
\label{sec:proofofamax}

This appendix contains no new material, but rather is intended as a
self-contained review of the proof of $a$-maximization, which
originally appeared in \cite{Intriligator:2003jj}, based on results
from \cite{Osborn:1998qu}.

The claim that $R$ locally maximizes
\beq
a(R_t) = \frac{3}{32}[3\Tr(R_t^3) - \Tr(R_t)],
\eeq
where $R_t = R_0 + \sum_I s_I F_I$ is equivalent to the statements
\beq
\label{eq:firstderivative}
\bullet&&\frac {32}{3}\frac{\ptl a}{\ptl s_I} \spceq 9\Tr(R R F_I)-\Tr(F_I)\spceq 0\quad\mbox{for each $F_I$.}
\\
\label{eq:secondderivative}
\bullet&&\frac {32}{3}\frac{\ptl^2 a}{\ptl s_I \ptl s_J}\spceq 18\Tr(R F_I F_J)\quad \mbox{is a negative-definite matrix}.
\eeq
Their proofs are essentially independent, and we will show each in turn.

\subsection{Proof of \Eq{eq:firstderivative}}

\Eq{eq:firstderivative} is a relation between the mixed anomaly
$\<\ptl^{\rho} J_{\rho}^I\, J^R_{\mu}\, J^R_{\nu}\>\propto\Tr(F_IRR)$
and the gravitational anomaly $\<\ptl^{\rho} J_{\rho}^I\,
T_{\mu\sigma}\,T_{\nu\gamma}\>\propto\Tr(F_I)$.  In a supersymmetric
theory, the flavor currents $J^{\mu}_I$ live in vector superfields
$J_I(z)$ with
\beq
J_I^{\mu}(x)=\s^{\mu}_{\a\dot\a}[\nabla^\a,\bar\nabla^{\dot\a}]J_I(z)|_{\th=0},
\eeq
where $z$ is a superspace coordinate standing for $(x,\th,\bar \th)$.
Meanwhile the $R$-current $J^R_{\mu}$, the stress tensor $T_{\mu\nu}$,
and the supersymmetry currents $\{S_{\mu}^\a,\bar S_{\mu \dot\a}\}$,
are components of a single ``supercurrent,"~\cite{Ferrara:1974pz}
\beq
T_{\mu}(z) &=& J_\mu^R(x)+\th^\a\p{S_{\mu\a}+\frac 1 3 (\s_\mu \bar \s^\rho S_\rho)_\a}+\bar\th_{\dot\a}\p{\bar S^{\dot\a}_\mu+\frac 1 3 \e^{\dot\a\dot\b}(\bar S_\rho \bar \s^\rho \s_\mu)_{\dot\b}}\\
&& + (\th\s^\nu \bar\th)\p{2 T_{\mu\nu}-\frac 2 3 \eta_{\mu\nu} T+\frac 1 4 \e_{\mu\nu\sigma\rho}\ptl^{[\rho}J_R^{\sigma]}}+\dots
\eeq
The supercurrent satisfies the conservation law
\beq
\label{eq:Tconservation}
\bar\nabla^{\dot \a}T_{\a\dot\a}=\nabla_\a L_T,
\eeq
where $L_T$ (a chiral superfield) is the trace anomaly, equal to the
variation of the action with respect to the chiral compensator in
supergravity.  In a conformal theory on a flat geometry with no
background fields, $L_T$ vanishes.

Notice that both correlation functions $\<J^I_{\rho} J^R_{\mu}
J^R_{\nu}\>$ and $\<J^I_{\rho} T_{\mu\sigma}T_{\nu\gamma}\>$ occur as
components of the superfield correlator
\beq
\label{threeptcorrleator}
\<T_{\mu}(z_1) T_{\nu}(z_2) J_I(z_3) \>.
\eeq
It follows that both anomalies $\Tr(RRF_I)$ and $\Tr(F_I)$ are
determined by this correlator.  $\<T_{\mu} T_{\nu} J_I\>$ will
generically have singularities as the points $z_i$ approach each
other.  This isn't a problem in an isolated CFT.  However, if we try
to couple the theory to background fields $\{E^{\mu}, A^I\}$ via a
coupling $\de \mathcal L = \int d^4\th (E^{\mu} T_{\mu} + A^I J_I)$,
then non-integrable singularities in correlators give rise to
divergences in the perturbation expansion $\<\exp(i\int
d^4x\,\de\mathcal L)\>$.  To make sense of the theory in non-trivial
backgrounds, we need to regularize the non-integrable singularities.
(For the asymptotically-free theories we consider in this paper, the
regularization comes about physically from the fact that the CFT
emerges from a different UV theory where the correlators are
better-behaved at short distances.)  Regularization introduces
anomalous contact terms into the flavor-current conservation law,
\beq
\bar D_3^2\<T_{\mu}(z_1) T_{\nu}(z_2) J_I(z_3) \> &=&\mbox{contact terms}.
\eeq
And we can read off the anomaly coefficients $\Tr(RRF_I)$ and
$\Tr(F_I)$ from different $\th$-components of these contact terms.

So far, these considerations have been true in any supersymmetric
theory.  However the enlarged symmetry group of a SCFT imposes
additional constraints.  In particular, $J_I$ and $T_{\mu}$ become
primary operators and their three-point function $\<T_{\mu}(z_2)
T_{\nu}(z_3) J_I(z_1) \>$ is uniquely determined up to an overall
constant by superconformal symmetry \cite{Osborn:1998qu}.  This
immediately implies that $\Tr(RRF_I)$ and $\Tr(F_I)$ are proportional
to each other, with a constant that's universal for any SCFT.  We can
fix this constant by examining the special case of a free chiral
superfield, which has $R$-charge $\frac 2 3$ and a single $U(1)$
flavor symmetry $F$.  In this case, $\Tr(RRF)=\Tr((\frac 2 3 -1)^2
F)=\frac 1 9 \Tr(F)$, which suffices to establish
\Eq{eq:firstderivative}.

\subsubsection{\texorpdfstring{Determining $\<T_{\mu}T_{\nu} J_I\>$ from Superconformal Invariance}{Determining <TTJ> from Superconformal Invariance}}

Since many readers may be unfamiliar with the results of
\cite{Osborn:1998qu}, we'd like to go into greater detail about how
superconformal symmetry fixes $\<T_{\mu} T_{\nu} J_I\>$, and in turn
the relation between $\Tr(RRF_I)$ and $\Tr(F_I)$.\footnote{Though we
  will directly follow calculations in \cite{Osborn:1998qu}, we will
  not make use of some of the more specialized notation, in the
  interest of making this section as accessible as possible.}

Superconformal primary operators $\cO$ are characterized by their
spin, and weights $(q,\bar q)$ such that $\dim(\cO)=q+\bar q$ and
$R(\cO)=\frac 2 3(q-\bar q)$.  The supercurrent $T_{\mu}$ has $q=\bar
q = \frac 3 2$, while the flavor currents $J_I$ have $q=\bar q = 1$.
These data completely determine an operator's transformation
properties under superconformal transformations, along with all
two-point functions (up to constants).  In particular for the flavor
currents, we have
\beq
\label{eq:twopointsupercorrelator}
\<J_I(z)J_J(0)\> &=& \frac {c_{IJ}} {x_+^2x_-^2},
\eeq
where $x_{\pm}$ are chiral and anti-chiral coordinates $x_\pm^\mu =
x^\mu\pm i\th\s^\mu\bar\th$, and $c_{IJ}$ are constants.  Let us pick a
basis such that $c_{IJ}=\de_{IJ}$, and focus on a single flavor
current $J_I=J$.

The superconformal group is generated by super-Poincare
transformations and inversions which act on superspace as
\beq
x_{\mp}'=\frac{x_\pm}{x_\pm^2},\quad
\th' = -i \frac{(x_{-}\.\s)\bar\th}{x_-^2},\quad
\bar\th' = i\frac{\th(x_+\.\s)}{x_+^2}.
\eeq
Under an inversion $z\to z'$, $J$ and $T_{\mu}$ transform as
\beq
J(z) &\to& J'(z)=\frac 1 {x_+^2 x_-^2} J(z')\\
T_{\mu}(z) &\to& T'_{\mu}(z)=\frac{I_{\mu\nu}(z)}{(x_+^2x_-^2)^{3/2}}T^{\nu}(z'),
\eeq
where
\beq
I_{\mu\nu}(z) &=& \frac{\tr(\s_{\mu} \bar\s_{\rho} \s_{\nu} \bar\s_{\sigma})x_-^{\rho}x_+^{\sigma}}{2(x_-^2x_+^2)^{1/2}}.
\eeq

As is perhaps familiar from the bosonic case, three-point functions
can be constructed from two-point functions together with OPE
relations.  For the correlator (\ref{threeptcorrleator}), this works
as follows.  By (super)translation invariance, it suffices to find
$\<T_{\mu}(z_1)T_{\nu}(z_2)J(0)\>$.  We will compute this as
$\lim_{z_3\to 0}\<T_{\mu}(z_1)T_{\nu}(z_2)J(z_3)\>$.  Consider an
inversion around $0$, taking $z_i\to z_i'$ in the correlator
\beq
\label{eq:doinversion}
\lim_{z_3\to 0}\<T_{\mu}(z_1)T_{\nu}(z_2)J(z_3)\>
&=&
\lim_{z_3\to 0}\frac{{I_{\mu}}^{\rho}(z_1){I_{\nu}}^{\sigma}(z_2)}{(x_{1+}^2x_{1-}^2x_{2+}^2x_{2-}^2)^{3/2}}\<T_{\rho}(z_1')T_{\sigma}(z_2')J(z_3')\>\frac{1}{x_{3+}^2x_{3-}^2}.
\eeq
Now the point $z_3'$ is approaching $\infty$ while $z_1'$ and $z_2'$
remain bounded, so we can safely use the $TT$ operator product
expansion,
\beq
\label{eq:TTOPE}
T_{\mu}(z_1')T_{\nu}(z_2')\sim \ldots+t_{\mu\nu}(z_1',z_2') J(z_2')+\ldots
\eeq
where ``$\dots$" represents other operators.  We'll see shortly that
the $J(z_2')$ term above is the only one that can contribute to the
correlator (\ref{threeptcorrleator}).  Note for the moment that the
only operators that have a nonvanishing two-point function with $J$
are $J$ and its descendants (obtained by acting on $J$ with momentum
and supersymmetry generators).

Using the OPE (\ref{eq:TTOPE}), the right-hand side of
(\ref{eq:doinversion}) becomes
\beq
\label{eq:invertedcorrelator}
\frac{{I_{\mu}}^{\rho}(z_1){I_{\nu}}^{\sigma}(z_2)}{(x_{1+}^2x_{1-}^2x_{2+}^2x_{2-}^2)^{3/2}}t_{\rho\sigma}(z_1',z_2')
\lim_{z_3\to 0}\<J(z_2')J(z_3')\>\frac{1}{x_{3+}^2x_{3-}^2}+\dots
\eeq
where ``$\dots$" now represents the contribution from descendants of
$J$.  As $z_3'\to\infty$, the factor $(x_{3+}^2x_{3-}^2)^{-1}$ grows
like $x_3'^4$, while the two-point function $\<J(z_2')J(z_3')\>$ dies
like $x_3'^{-4}$.  By contrast, $\<\cO(z_2')J(z_3')\>$ dies faster
than $x_3'^{-4}$ whenever $\cO$ is a descendant of $J$.  Hence,
contributions from such operators don't survive in the limit
$z_3'\to\infty$.  In fact, since shifting $J(z_2')\to J(z_2'+\de z)$
changes $J(z_2')$ by descendants, we can freely replace $J(z_2')$ with
$J(0)$ in the limit, so the last two factors in
(\ref{eq:invertedcorrelator}) become
\beq
\lim_{z_3\to 0}\<J(z_2')J(z_3')\>\frac{1}{x_{3+}^2x_{3-}^2}&=&\lim_{z_3\to0}\<J(0)J(z_3')\>\frac{1}{x_{3+}^2x_{3-}^2} \spceq 1.
\eeq
This leaves us with
\beq
\<T_{\mu}(z_1)T_{\nu}(z_2)J(0)\>
&=&
\frac{{I_{\mu}}^{\rho}(z_1){I_{\nu}}^{\sigma}(z_2)}{(x_{1+}^2x_{1-}^2x_{2+}^2x_{2-}^2)^{3/2}}t_{\rho\sigma}(z_1',z_2').
\eeq

Finally, we will show that the OPE coefficient $t_{\mu\nu}$ is
determined up to a constant by symmetry and conservation of the
supercurrent.  By supertranslation invariance, we may compute
$t_{\mu\nu}(z)\equiv t_{\mu\nu}(z,0)$.  Exchanging
$T_{\mu}\leftrightarrow T_{\nu}$ gives $t_{\mu\nu}(z)=t_{\nu\mu}(-z)$.
Also, $t_{\mu\nu}$ is real and thus a function of $x^{\mu}$ and
$p^{\mu}\equiv \th\s^{\mu}\bar\th$ alone.  Since $p_{\mu}
p_{\nu}=\frac 1 4\eta_{\mu\nu}p^2$ and $p^2p_{\mu}=0$, we see that the
most general form of $t_{\mu\nu}$ with the right scaling dimension and
symmetry properties is
\beq
\label{eq:generalOPEcoeff}
t_{\mu\nu}(z)&=&\frac{\eta_{\mu\nu}}{x^4}\p{A+B\frac{p^2}{x^2}}+\frac{x_{\mu}x_{\nu}}{x^6}\p{C+D\frac{p^2}{x^2}}+\frac{E}{x^6}\e_{\mu\nu\rho\sigma}x^{\rho}p^{\sigma},
\eeq
where $A,B,C,D,E$ are real constants.

The conservation law \Eq{eq:Tconservation} implies $\bar
D^{\dot\a}t_{\a\dot\a \b\dot\b}=0$, or equivalently
\beq
\th \s^{\rho}\bar\s^{\mu}\p{\frac{\ptl}{\ptl p^{\rho}}-i\frac{\ptl}{\ptl x^{\rho}}}t_{\mu\nu} &=& 0.
\eeq
Inserting the general form (\ref{eq:generalOPEcoeff}) gives $C=-4A$
and $B=D=E=0$, which completes the proof that $\<T_{\mu}T_{\nu}J\>$ is
determined up to an overall constant.

\subsection{Proof of \Eq{eq:secondderivative}}

The trace $\Tr(RF_IF_J)$ is proportional to the $R$-current anomaly
$\ptl_\mu J_R^\mu$ in the presence of background $U(1)$ flavor fields
$A_I$.  This is encoded in the supertrace $L_T\sim \Tr(W_I^\a
W_{J\a})$.  However, $L_T$ also encodes the trace of the stress-energy
tensor $\Theta=T^{\mu}_{\mu}$ \cite{Anselmi:1997am},
\beq
i\nabla_\mu J_R^\mu &=& \frac 1 2\{\nabla^\a,\bar\nabla^{\dot\a}\}T_{\a\dot\a}|_{\th=0}=(\nabla^2L_T-\bar\nabla^2\bar L_T)|_{\th=0}\\
\Theta &=& \frac{3}{8}[\nabla^\a,\bar\nabla^{\dot\a}]T_{\a\dot\a}|_{\th=0}=\frac 3 4(\nabla^2L_T+\bar\nabla^2 \bar L_T)|_{\th=0} .
\eeq
These relations fix $\Theta(x)$ in terms of the $R$-anomaly, a
familiar fact that we used, for instance to calculate the
$\SU(5)_\GUT$ $\b$-function in \Eq{eq:NSVZrewrite}.  Letting
$\<\.\>_A$ denote an expectation value in the flavor background, we have
\beq
\<\ptl_\mu J^\mu_R(x)\>_A &=& \Tr(RF_IF_J)\frac{1}{16\pi^2}F_I^{\mu\nu}\tl F_{J\mu\nu}(x)\\
\<\Theta(x)\>_A &=& -\Tr(RF_IF_J)\frac{3}{32\pi^2}F_I^{\mu\nu}F_{J\mu\nu}(x).
\eeq
So we can compute $\Tr(RF_IF_J)$ by examining how the $A_I$ break
scale invariance, leading to nonzero $\Theta$.

Conformal symmetry completely determines the two-point function of
flavor currents
\beq
\label{eq:twocurrentcorrelator}
\<J_I^\mu(x) J_J^\nu(0)\> &=& \frac{\tau_{IJ}}{(2\pi)^4}(\Box g^{\mu\nu}-\ptl^\mu\ptl^\nu)\frac{1}{x^4}
\eeq
where $\tau_{IJ}$ is a constant matrix which must be positive-definite
in a unitary theory ($\tau_{IJ}$ is related to $c_{IJ}$ in \Eq{eq:twopointsupercorrelator} by $c_{IJ}=\frac{\tau_{IJ}}{(4\pi)^4}$).
This has a non-integrable singularity at $x=0$,
which we must regulate in the presence of the flavor fields $A_I^\mu$.
If we introduce a cutoff at the scale $\frac 1 x \sim \L$, we need the
counterterm
\beq
\de\mathcal L &=& \frac{\tau_{IJ}}{32\pi^2}\log \frac{\mu}{\L}F_I^{\mu\nu}F_{J\mu\nu}
\eeq
to keep the physics fixed.  Thus we have
\beq
\left\<\int d^4x\, \Theta(x)\right\>_A &=& \mu\frac{d}{d\mu}\<1\>_A \spceq\left\<\int d^4 x\,\frac{\tau_{IJ}}{32\pi^2}F_I^{\mu\nu}F_{J\mu\nu}(x)\right\>_A,
\eeq
which implies $\Tr(RF_IF_J)=-\frac 1 3 \tau_{IJ}$.  Recall that
$\tau_{IJ}$ is positive-definite by unitarity, so we conclude that
$\Tr(RF_IF_J)$ is negative-definite.

\section{CFT Exit and Dual Descriptions}
\label{sec:CFTexit}

What happens when conformal symmetry is broken?  In this appendix we
will consider more carefully what happens when we deform our
vector-like models with mass terms. In general we will find that this
leads to non-trivial mass mixing between the fields $T_i$ and
composite states in the CFT, such that the Yukawa interactions among
light modes are modified by mixing angles.  The precise mixing angles
depend non-trivially on the mass deformations.  We will not be able to
compute them exactly, but we can say something about their structure
in certain limits.  Along the way, we'll learn something about Seiberg
duality in superconformal flavor models.

We will carry out much of the discussion in the general context of
$\SU(N_c)$ supersymmetric QCD with $N_f$ flavors $\{Q,\bar Q\}$
coupled to singlets $T$, specializing to the toy model of
Section~\ref{sec:scfts} (where $Q=\{S,X\}$) when appropriate. In
order to determine the mixing angles mentioned above, we will need
to know the masses of fluctuations around the vacua of our theory
at the scale where conformal symmetry is broken. In some cases our
theory (considering just one generation for simplicity) can be
described an effective superpotential \beq \label{massmixingW} W
&=& m_1T_1\Phi_1+m_2 \Phi_2\Phi_1 , \eeq where
$\{T_1,\Phi_1,\Phi_2\}$ are canonically normalized fields.
Letting $\sin\a=m_2/\sqrt{m_1^2+m_2^2}$, we see that the linear
combination $T_1\cos\a+\Phi_2\sin\a$ becomes massive with
$\Phi_1$, leaving a canonically normalized massless mode
$T'=T_1\sin\a-\Phi_2\cos\a$ with the Yukawa interactions \beq W
&=& \sin^2\a\, \l^u_{11}T'T'H + \sin\a\, \l^u_{1j}T'T_j H +
\sin\a\, \l^d_{1j}T' \bar F_j \bar H. \eeq If the mixing angle is
$O(1)$, this will not be an important effect given that we have
already assumed arbitrary $O(1)$ couplings in the UV.  On the
other hand, if $\alpha$ is small then this mixing will behave like
an additional contribution to $\epsilon_{T_1}$. In the main body
of this text we have implicitly assumed the former rather than the
latter.  Here we will try to explore the extent to which this
assumption is justified.

In order to estimate these mixing effects, there are then two
important questions we should try to answer.  Firstly, what are
the appropriate degrees of freedom to describe fluctuations in the
exotic sector that couple to $T_1$ at or below the scale of
conformal symmetry breaking?  Secondly, what is the K\"ahler
potential for those degrees of freedom? The answer to the first
question depends on how strongly coupled the CFT is.  When $N_f$
is near $3N_c$, we should use the electric description in terms of
elementary quarks $\{Q,\bar Q\}$.  At the other end of the
conformal window where $N_f$ is near $\frac 3 2 N_c$, the magnetic
variables $\{M,q,\bar q\}$ are more appropriate.

As a first approximation, we will take the K\"ahler potential to
be determined only by the anomalous dimensions in the CFT, right
down to the scale of conformal symmetry breaking $\L_c$.  That is,
we will ignore the contribution of mass deformations to the
anomalous dimensions.  These effects are not calculable in a
strongly-coupled theory, but it is reasonable to make the
assumption that these ``threshold effects" are at most $O(1)$.  We
will also for the most part neglect RGE running below $\L_c$.  For
instance, if conformal symmetry is broken when some of the
electric quarks get mass $\L_c$ while others remain relatively
light, we will eschew the details of running and matching between
$\L_c$ and the lighter scale.  Ignoring these effects is
justifiable if the theory is weakly coupled below the scale $\L_c$
and there are no large logarithms, or if the remaining fields are
simply decoupled from the visible sector.

At a scale $\mu$, the electric theory has an effective
superpotential and gauge kinetic terms given by \beq
W_{electric}&=& h_* T Q\bar Q + \Tr(m(\mu)Q\bar Q) +
\frac{b}{16\pi^2} \log\p{\frac{\mu}{\L_h}} \Tr(W^\a W_\a), \eeq
where $T,Q,$ and $\bar Q$ are canonically normalized and we
suppress flavor indices. The coupling $h_*$ is at its fixed-point
value, while the mass matrix is given by \beq
m(\mu)=m_{UV}\p{\frac \mu \L}^{\frac 3 2 R(Q\bar Q)-2}. \eeq
Finally, $\L_h$ is the holomorphic scale\footnote{ It may seem
unusual that a CFT has a distinguished scale.  However, this is
just an artifact of our choice of normalization for the quarks,
and doesn't reflect breaking of conformal invariance. $\L_h$ is
the position of the ostensible Landau pole if the gauge coupling
were to undergo na\"ive one-loop running from the scale $\mu$.
But the anomalous dimensions of the quarks modify the running of
$g$, and ensure that this pole is unphysical.  If we flow to a
different energy scale $\mu'$, we should rescale the quarks
$Q,\bar Q$ to canonically normalize them.  However, the rescaling
anomaly then shifts $\L_h\to\L_h'= \frac{\mu'}{\mu}\L_h$.} \beq
\L_h &=& \mu \exp\p{\frac{2\pi i\tau_*}{b}} \ \ =\ \ \mu
\exp\p{-\frac{8\pi^2}{bg_*^2}+i\frac{\th}{b}}, \eeq where
$b=3N_c-N_f$ and $g_*$ is the fixed-point value of the gauge
coupling.

If our CFT is weakly coupled ($N_c g_*^2 / 8\pi^2 \ll 1$), then
the electric quarks $Q,\bar Q$ are good degrees of freedom.  The
mass term becomes important when $m(\mu)\sim \mu$, at which point
$Q$ and $\bar Q$ are lifted from the spectrum.  It's clear that
perturbatively there is no mass mixing between $T$ and exotic
states, and the primary effect of integrating out the quarks at
$\mu=\L_c$ is to correct the K\"ahler potential \beq
\label{Kahlercorrection} T^\dag T &\to & T^\dag T\p{1+\frac{c
|h_*|^2}{16\pi^2}+\dots} \eeq and introduce higher-dimension
operators suppressed by $\L_c$.

By contrast, when the electric theory is very strongly coupled, a
more appropriate description of the degrees of freedom is in terms
of the dual magnetic variables $\{M,q,\bar q\}$ with gauge group
$\SU(\tl N)=\SU(N_f-N_c)$~\cite{Seiberg:1994pq}.  When $T$, $M$,
$q$, and $\bar q$ are canonically normalized at the scale $\mu$,
the magnetic theory has an effective superpotential and gauge
kinetic terms given by \beq W_{magnetic} &=& \tl h_* \mu TM +
\l_* M q\bar q + \mu\Tr(\tl m(\mu) M) + \frac{\tl
b}{16\pi^2}\log\p{\frac{\mu}{\tl \L_h}}\Tr(W^\a W_\a), \eeq where
$\tl h_*$ and the block matrix $\l_*$ are at their fixed-point values, \beq \tl
\L_h &=& \mu \exp\p{\frac{2\pi i \tl\tau_*}{\tl b}} \ \ =\ \ \mu
\exp\p{-\frac{8\pi^2}{\tl b \tl g_*^2} + i\frac{\tl\th}{\tl b}}
\eeq is the magnetic holomorphic scale, and $\tl b=3\tl N-N_f$.
Since the coupling between $T$ and $M$ is simply a mass term in
this description, one approach would be to simply integrate out
these massive fields.  However, expanding the theory around its
supersymmetric vacua will in general induce additional mass terms
for the mesons $M$, leading to the mixing effects discussed above.
In order to better understand these mixing effects we will for now
leave $T$ in the description of the theory until we know which
linear combination of fields becomes massive.

A partial dictionary between electric and magnetic variables follows
from matching vacuum superpotentials.  We will here only
consider singlet couplings that lead to vacua where the VEV of $T$
(as well as the meson it couples to) vanishes, since otherwise the
GUT group would be broken in a problematic way.  We will also assume
that $m(\mu)$ is of maximal rank.  Since we only care about holomorphic
information for the moment, we can be brazen about integrating out degrees of
freedom without worrying about their physical masses.

In the electric theory, integrating out the quarks leaves pure
$\SU(N_c)$ SYM with scale $\L_L=(\det m(\mu)\L_h^b)^{1/3N_c}$,
which has $N_c$ vacua with superpotential \beq W_{vac}&=&
N_c\L_L^3 \ \ =\ \ N_c(\det m(\mu)\L_h^b)^{1/N_c}. \eeq Meanwhile
in the magnetic theory, the dual quarks have a meson-dependent
mass matrix $\l_* M$.  Integrating them out gives pure $\SU(\tl
N)$ SYM with scale $\tl \L_L = (\det (\l_* M) \tl \L_h^{\tl
b})^{1/3\tl N}$, which confines and leaves us with an effective
superpotential \beq W_{eff}(M) &=& \tl h_* \mu TM+ \mu \Tr(\tl
m(\mu) M)+\tl N(\det (\l_* M) \tl \L_h^{\tl b})^{1/\tl N}. \eeq
Finally, extremizing $W_{eff}$, we find \beq \label{mesonvev}
\<M\> &=& \frac 1 \mu \p{\frac{(-1)^{\tl N} \det(\mu \tl m(\mu)\l_*^{-1})}{\tl\L_h^{\tl b}}}^{1/N_c} {\tl m(\mu)}^{-1}\\
W_{eff}(\<M\>) &=& N_c \p{\frac{(-1)^{\tl N} \det(\mu \tl
m(\mu)\l_*^{-1})}{\tl\L_h^{\tl b}}}^{1/N_c}. \eeq Matching
$W_{eff}(\<M\>)=W_{vac}$ then implies the relation \beq \tl m(\mu)
\spceq m(\mu) \frac{\l_* \hat \L}{\mu}, \qquad
\eeq
where $\hat \L^{N_f} = (-1)^{\tl N}\L_h^b \tl \L_h^{\tl
b}=\mu^{N_f} e^{2\pi i(\tau_*+\tl \tau_*+\tl N/2)}$.

Holomorphy only tells us about expectation values of chiral
operators.  If we're interested in physical degrees of freedom and
masses, the correct picture of the magnetic vacua depends more
intricately on the structure of the mass matrix $\tl m(\mu)$.  For
example, let us start by supposing that $\tl m(\mu)$ has a single
large eigenvalue $\tl m_1$ at the scale $\mu = \L_c$, where $\L_c
\sim \tl m_1$ is the scale at which the CFT description breaks down.
 Ignoring the other eigenvalues for the moment, the equation of
motion $\l_* q_1\bar q_1 = \L_c \tl m_1$ implies that $q_1$ and
$\bar q_1$ get VEVs, Higgsing the dual gauge group
$\SU(\tl N)\to \SU(\tl N-1)$.\footnote{This type of analysis is
only trustworthy when $\tl m(\mu)$ has at most $N_c-1$ large
eigenvalues. For a general rank-$N_f$ mass matrix, the equation of
motion $\l_* q\bar q = \L_c \tl m$ cannot be satisfied
classically, since $\mathrm{rank}(q\bar q)\leq N_c$. This ``rank
condition" is violated non-perturbatively in the supersymmetric
vacua, so there is no contradiction with \Eq{mesonvev}.  In
free magnetic theories, the rank condition implies the existence
of metastable SUSY-breaking vacua \cite{Intriligator:2006dd}.  In
this paper, we will only consider supersymmetric vacua.  However,
it would be interesting to think about metastable SUSY breaking in
superconformal flavor models, perhaps along the lines of~\cite{Izawa:2009nz}.}
At the same time, the VEVs
$\<q_1\>,\<\bar q_1\>$ lift some of the meson flavors via mass
terms \beq W = \l_* \<q_1\> \bar q_i M_{1i} + \l_* \<\bar q_1\>
q_i M_{i1}+ \dots . \eeq  Of course, the remaining meson flavors
could remain somewhat light, but this might not be important for
us.

Now, suppose in our toy model we take $\tl m_S \gg \tl m_X$. Then
the GUT singlet dual quarks $\{s,\bar s\}$ get VEVs
$\sim\sqrt{\L_c\, \tl m_S/\l_*}$, yielding an effective
superpotential \beq W &=& \tl h_* \L_c\, T_1 M_{S\,\bar{X}} +
\sqrt{\l_* \L_c\, \tl m_S}\, \bar x M_{S\,\bar{X}} + \dots, \eeq
where $\bar x$ is a dual quark in the $\tenrep$ representation of
$\SU(5)_{\GUT}$.  We see here that $T_1$ mixes with $\bar x$, and
the mixing angle $\sin \a$ depends on the ratio of $\sqrt\l_*$ and
$\tl h_*$, as well as the precise relation between $\L_c$ and $\tl
m_1$ and any threshold corrections in the K\"ahler potential.

Next let us consider the different limit in which all the masses
are the same at $\mu=\L_c$, i.e. $\tl m_{ij}(\L_c)=\tl
m\,\de_{ij}$. In the limit where the magnetic theory is
weakly-coupled, we can verify that the dual quarks are much
heavier than the mesons, and should be integrated out.  In this
case the dual quark masses set the scale of conformal symmetry
breaking, $\L_c \sim \l_*\<M\>$.  Solving this equation for $\tl
m$ then yields \beq \tl m &\sim& -\l_*(\L_c^{N_c-\tl N}\tl
\L_h^{\tl b})^{1/\tl N} \ \ \sim \ \ -\l_* \L_c e^{2\pi i \tl
\tau_*/\tl N} \ \ \ll\ \ \L_c, \eeq where the last relation holds
(up to the phase) when $\tl g_*^2 \tl N / 8\pi^2 \ll 1$ due to the
exponential suppression.  Further, we may estimate
\beq\label{mesonsgood} \left.\frac{\ptl^2 W_{eff}}{\ptl
M^2}\right|_{M=\<M\>} &\sim& \frac{\tl m \mu}{\<M\>} \ \ \sim \ \
\l_* \tl m \ \ \ll\ \ \L_c , \eeq showing it is indeed physically
sensible in this limit to integrate out $q,\bar q$, leaving us
with an effective theory of mesons with superpotential
$W_{eff}(M)$.

Specializing to our toy model, the superpotential after conformal
symmetry breaking is \beq W_{eff}(\<M\>+\hat M) &=& \tl h_* \L_c
\,T_1 \hat M_{S\,\bar{X}} + \l_* \tl m \,\hat M_{X\,\bar{S}}\hat
M_{S\,\bar{X}}+\dots, \eeq where $\hat M$ are fluctuations around
the VEV $\<M\>$.  Note that integrating out $q,\bar q$ will give a
correction to the K\"ahler potential of $\hat M$ similar to
\Eq{Kahlercorrection}.  However, in the weakly-coupled
magnetic limit, $\hat M$ is close to canonically normalized at
$\mu=\L_c$.  This superpotential is then of the general form
(\ref{massmixingW}), so we find mass mixing between $T_1$ and
$\hat M_{X\,\bar{S}}$, with an angle depending on the fixed-point
values of the couplings $\tl h_*, \l_*$, and $\tl\tau_*$.  In the
very weakly-coupled limit, $\tl m / \L_c$ is suppressed as
$e^{-8\pi^2/\tl N\tl g_*^2}$ and hence we expect the mixing angle
$\sin\a \ll 1$.  On the other hand, it is important to note that
approaching this weakly-coupled (Banks-Zaks)
limit~\cite{Banks:1981nn,Kogan:1995mr} requires being at very large $N$, and none of the models considered in the
present work are at sufficiently large $N$ so as to really approach
this limit.

Finally, we would like to comment on the alternative approach of
simply integrating out the $\{T,M\}$ pair at the {\it top} of the
conformal window.  Doing so and running down to arbitrary $\mu$
generates operators such as
\beq
W = -
\frac{\l_*}{\tl h_* \mu} \left(\frac{\mu}{\L}\right)^{\frac32 R(q
\bar{q}) - 1} y^d_{1 j} (q\bar{q}) \bar{F}_j \bar{H} + \dots,
\eeq
along with similar operators corresponding to the other Yukawa interactions.
Note that these operators are generated at the scale $\L$ with the na\"ive
$1 / \L$ suppression, but are then anomalously affected by the CFT
dynamics at lower energies.  In the case that the singlet dual
quark $s$ acquires a VEV, these couplings lead to the same
predictions as before after identifying the dual quark $\bar{x}$
with the low-energy $\tenrep$.  On the other hand, when the mesons
acquire VEVs and the non-perturbative superpotential is important,
the situation is slightly more subtle.  In this case, the dual
quark mass matrix is now a linear combination of $M$ and $\bar{F}_j
\bar{H}$, and this linear combination will enter the non-perturbative
superpotential.  After expanding the non-perturbative superpotential around
$\<M\>$, the Yukawa couplings are then recovered by identifying $M_{X \bar{S}}$
with the low-energy $\tenrep$.  Similar considerations will apply to the
$y^u_{1 j}$ couplings for $j \neq 1$.  Somewhat more care is required to understand
how the operator $y^u_{1 1}(q\bar{q})(q\bar{q})H$ affects the non-perturbative
superpotential when the dual quarks are integrated out.  However, this could for
example be obtained by using chiral ring relations to determine an effective glueball
superpotential, which when minimized and expanded around $\<M\>$ should recover
the same predictions as before.

To summarize, here we've identified at least three kinds of behavior when
conformal symmetry is broken, depending on how strongly-coupled
the conformal sector is and the structure of the mass matrix
$m(\mu)$ (or equivalently $\tl m(\mu)$).  When the electric theory is weakly-coupled, exotic states
are lifted and the K\"ahler potential of $T_1$ receives small corrections.
When the magnetic theory is weakly coupled, $T_1$ can mix with either meson
fluctuations or dual quarks, with different mixing angles
depending on the structure of $\tl m(\mu)$.  We expect a variety of
mixing angles are possible away from these special cases.  In this
paper, we have assumed that RGE running gives the dominant
contribution to the Yukawa hierarchies.  This simplifies our
analysis, but it is not guaranteed to be what nature chooses.
Some models that we rule out as {\it pure} superconformal flavor
models may be salvageable if the Yukawa hierarchies come from a
hybrid of RGE running and mixing angles.

\section{Models with Two $3$-field or $4$-field $T_i$ Couplings}
\label{apx:listofmodels}

In this appendix, we list all vector-like exotic sectors with a
simple, non-IR free gauge group and two
representation-theoretically distinct three-field or four-field
couplings to $T_i$'s.  In scanning over models, we have been aided by \cite{LiEComputerAlgebra}.
In many cases, these exotic sectors can be
extended by adding GUT singlets, for instance to get more
couplings to SM $\fivebarrep$'s.  However, adding more matter
makes the CFT less strongly coupled, and typically reduces the
anomalous dimensions of $T_1$ and $T_2$. Regardless, here we only
display the fields directly involved in giving anomalous
dimensions to $T_1$ and $T_2$.

The condition that the two couplings $T_1\cO_1$ and $T_2\cO_2$ be
representation-theoretically distinct is an elegant way to ensure
that no symmetry forces $\g_{T_1}=\g_{T_2}$.  Another strategry is
to allow the two couplings to be representation-theoretically
identical, but break the symmetry between them with other
superpotential deformations.  This is the solution used, for
instance, in the model of
Section~\ref{sec:incalculable} (originally
from~\cite{Nelson:2000sn}).  However, we won't attempt to classify
such models here.

In addition to the representation content of a superconformal flavor
model, one has to decide which relevant deformations to include in the
superpotential.  We have not attempted to study all possible
deformations here.  Rather, we chose to focus in the main text on the
phenomenologically interesting deformations of the most promising models.

\pagebreak

\subsection{\texorpdfstring{Models without $\tenrep$'s}{Models without 10's}}

\begin{table}[h!]
\begin{center}
\begin{tabular}{ | l | c | c | }
\hline
                    & $\SU(5)_\GUT$ & $\SU(N)$ \\
\hline
$Q_1 + \bar{Q}_2$   & $\fiverep + \fivebarrep$      & $\Yfund$ \\
$\bar{Q}_1 + Q_2$   & $\fivebarrep + \fiverep$      & $\Yafund$ \\
$A$               & $\trivrep$              & Ad. \\
\hline
\end{tabular}
\end{center}
\caption{Couplings $T_1\bar Q_1\bar Q_2+T_2\bar Q_1A\bar Q_2$.}
\end{table}

\begin{table}[h!]
\begin{center}
\begin{tabular}{ | l | c | c | }
\hline
                    & $\SU(5)_\GUT$ & $\SO(N)$ \\
\hline
$Q_{1,2}+\bar{Q}_{1,2}$   & $2\times(\fiverep+\fivebarrep)$      & $\Yfund$ \\
$A$               & $1$              & $\Yasymm$ or $\Ysymm$ \\
\hline
\end{tabular}
\end{center}
\caption{Couplings $T_1\bar Q_1\bar Q_2+T_2\bar Q_1A\bar Q_2$.}
\end{table}

\begin{table}[h!]
\begin{center}
\begin{tabular}{ | l | c | c | }
\hline
                    & $\SU(5)_\GUT$ & $\Sp(2N)$ \\
\hline
$Q+\bar{Q}$       & $\fiverep + \fivebarrep$      & $\Yfund$ \\
$A$               & $\trivrep$              & $\Yasymm$ \\
\hline
\end{tabular}
\end{center}
\caption{Couplings $T_1 \bar Q\bar Q+T_2 \bar Q A \bar Q$.}
\end{table}

\begin{table}[h!]
\begin{center}
\begin{tabular}{ | l | c | c | }
\hline
                    & $\SU(5)_\GUT$ & $\SU(3)$ \\
\hline
$Q+S$       & $\fiverep + \trivrep$ & $\Yfund$ \\
$\bar Q + \bar S$ & $\fivebarrep + \trivrep$ & $\Yafund$ \\
\hline
\end{tabular}
\end{center}
\caption{Couplings $T_1 Q^3+T_2 \bar S \bar Q^2$}
\end{table}

\begin{table}[h!]
\begin{center}
\begin{tabular}{ | l | c | c | }
\hline
                    & $\SU(5)_\GUT$ & $G_2$ \\
\hline
$Q + \bar Q + S$       & $\fiverep + \fivebarrep + \trivrep$      & $\Yfund$ \\
\hline
\end{tabular}
\end{center}
\caption{Couplings $T_1 Q^3+T_2 S\bar Q^2$.}
\end{table}

\pagebreak

\subsection{\texorpdfstring{Models with $\tenrep$'s}{Models with 10's}}

\begin{table}[h!]
\begin{center}
\begin{tabular}{ | l | c | c | }
\hline
                    & $\SU(5)_\GUT$ & $\SU(N)$ \\
\hline
$X+\bar{Q}+S$       & $\tenrep + \fivebarrep + \trivrep$ & $\Yfund$ \\
$\bar{X}+Q+\bar{S}$ & $\tenbarrep + \fiverep + \trivrep$ & $\Yafund$ \\
\hline
\end{tabular}
\end{center}
\caption{Couplings $T_1\bar X S+T_2XQ$.}
\end{table}

\begin{table}[h!]
\begin{center}
\begin{tabular}{ | l | c | c | }
\hline
                    & $\SU(5)_\GUT$ & $\SU(N)$ \\
\hline
$X+Q_1+\bar{Q}_2$       & $\tenrep + \fiverep + \fivebarrep$ & $\Yfund$ \\
$\bar{X}+\bar Q_1+Q_2$ & $\tenbarrep + \fivebarrep + \fiverep$ & $\Yafund$ \\
\hline
\end{tabular}
\end{center}
\caption{Couplings $T_1\bar Q_1\bar Q_2+T_2 X Q_2$.}
\end{table}

\begin{table}[h!]
\begin{center}
\begin{tabular}{ | l | c | c | }
\hline
                    & $\SU(5)_\GUT$ & $\SU(N)$ \\
\hline
$X+S$       & $\tenrep + \trivrep$ & $\Yfund$ \\
$\bar{X}+\bar{S}$ & $\tenbarrep + \trivrep$ & $\Yafund$ \\
$A$ & $\trivrep$ & Ad.\\
\hline
\end{tabular}
\end{center}
\caption{Couplings $T_1\bar X S+T_2\bar X A S$.}
\end{table}

\begin{table}[h!]
\begin{center}
\begin{tabular}{ | l | c | c | }
\hline
                    & $\SU(5)_\GUT$ & $\SU(N)$ \\
\hline
$X+S$       & $\tenrep + \trivrep$ & $\Yfund$ \\
$\bar{X}+\bar{S}$ & $\tenbarrep + \trivrep$ & $\Yafund$ \\
$B+\bar B$ & $\trivrep$ & $\left(\Yasymm + \overline{\Yasymm}\right)$ or $\left(\Ysymm + \overline{\Ysymm}\right)$\\
\hline
\end{tabular}
\end{center}
\caption{Couplings $T_1\bar X S+T_2\bar XB\bar S$.}
\end{table}

\begin{table}[h!]
\begin{center}
\begin{tabular}{ | l | c | c | }
\hline
                    & $\SU(5)_\GUT$ & $\SU(N)$ \\
\hline
$X+Q$       & $\tenrep + \fiverep$ & $\Yfund$ \\
$\bar{X}+\bar{Q}$ & $\tenbarrep + \fivebarrep$ & $\Yafund$ \\
$B+\bar B$ & $\trivrep$ & $\Yasymm + \overline{\Yasymm}$\\
\hline
\end{tabular}
\end{center}
\caption{Couplings $T_1X \bar B Q+T_2\bar Q B\bar Q$.}
\end{table}

\begin{table}[h!]
\begin{center}
\begin{tabular}{ | l | c | c | }
\hline
                    & $\SU(5)_\GUT$ & $\SU(N)$ \\
\hline
$X+\bar Q$       & $\tenrep + \fivebarrep$ & $\Yfund$ \\
$\bar{X}+Q$ & $\tenbarrep + \fiverep$ & $\Yafund$ \\
$B+\bar B$ & $\trivrep$ & $\Yasymm + \overline{\Yasymm}$\\
\hline
\end{tabular}
\end{center}
\caption{Couplings $T_1X Q+T_2\bar Q B \bar Q$.}
\end{table}

\pagebreak

\begin{table}[h!]
\begin{center}
\begin{tabular}{ | l | c | c | }
\hline
                    & $\SU(5)_\GUT$ & $\SU(N)$ \\
\hline
$X+\bar Q$       & $\tenrep + \fivebarrep$ & $\Yfund$ \\
$\bar{X}+Q$ & $\tenbarrep + \fiverep$ & $\Yafund$ \\
$A$ & $\trivrep$ & Ad.\\
\hline
\end{tabular}
\end{center}
\caption{Couplings $T_1XQ+T_2XAQ$.}
\end{table}

\begin{table}[h!]
\begin{center}
\begin{tabular}{ | l | c | c | }
\hline
                    & $\SU(5)_\GUT$ & $\SO(N)$ \\
\hline
$X + \bar X + S$   & $\tenrep + \tenbarrep + \trivrep$      & $\Yfund$ \\
$A$               & $\trivrep$              & $\Yasymm$ or $\Ysymm$ \\
\hline
\end{tabular}
\end{center}
\caption{Couplings $T_1\bar XS+T_2\bar XAS$.}
\end{table}

\begin{table}[h!]
\begin{center}
\begin{tabular}{ | l | c | c | }
\hline
                    & $\SU(5)_\GUT$ & $\Sp(2N)$ \\
\hline
$X+\bar X+Q+\bar Q$       & $\tenrep + \tenbarrep + \fiverep + \fivebarrep$      & $\Yfund$ \\
\hline
\end{tabular}
\end{center}
\caption{Couplings $T_1 \bar Q\bar Q+T_2 XQ$.}
\end{table}

%%%%%%%%%%%%%%%%%%%%%%%%%%%%%%%%%%%%%%%%%%%%%%%%%%%%%%%%%%%%%%%%%

%\newpage

\end{document}